\documentclass[preprint]{aastex631}
\def\simgr{\,\hbox{\hbox{$ > $}\kern -0.8em \lower 1.0ex\hbox{$\sim$}}\,}
\def\simle{\,\hbox{\hbox{$ < $}\kern -0.8em \lower 1.0ex\hbox{$\sim$}}\,}

\shortauthors{Halpern \& Thorstensen}
\shorttitle{Hard X-ray Selected Cataclysmic Binaries}

\def\int{INTEGRAL/IBIS}
\def\chandra{Chandra}
\def\xmm{XMM-Newton}
\def\sw{Swift}

\newcommand{\swiftOhSevenOne}{Swift J0717.8$-$2156}
\newcommand{\swiftOhSevenFour}{Swift J0749.7$-$3218}
\newcommand{\swiftOhEight}{Swift J0820.6$-$2805}
\newcommand{\igrOneEightOh}{IGR J18017$-$3542}
\newcommand{\igr}{IGR J18151$-$1052}
\newcommand{\axj}{AX J1832.3$-$0840}
\newcommand{\pbcOneEight}{PBC J1841.1+0138}
\newcommand{\igrOneEightFour}{IGR J18434$-$0508}
\newcommand{\swiftOneNineOh}{Swift J1909.3+0124}
\newcommand{\swiftTwoTwo}{Swift J2237.2+6324}
\newcommand{\numberOfTargets}{ten}
\newcommand{\NumberOfTargets}{Ten}

\begin{document}

\title{Optical Studies of \NumberOfTargets\ Hard X-ray Selected Cataclysmic Binaries}

\author[0000-0003-4814-2377]{Jules P. Halpern}
\affiliation{Department of Astronomy, Columbia University, 550 West 120th Street, New York, NY 10027, USA}
\author[0000-0002-4964-4144]{John R. Thorstensen}
\affiliation{Department of Physics and Astronomy, Dartmouth College, Hanover NH 03755, USA}

\begin{abstract}
We conducted time-resolved optical spectroscopy and/or photometry
of \numberOfTargets\ cataclysmic binaries that were discovered
in hard X-ray surveys, with the goal of measuring their orbital
periods and searching for evidence that they are magnetic.
Four of the objects in this study are new optical identifications:
\igrOneEightOh, \pbcOneEight, \igrOneEightFour, and \swiftOneNineOh.
A 311.8~s, coherent optical pulsation is detected from \pbcOneEight,
as well as eclipses with a period of 0.221909~days.
A 152.49~s coherent period is detected from \igrOneEightFour. 
A probable period of 389~s is seen in \igr, in agreement with a known X-ray spin period.
We also detect a period of 803.5~s in an archival X-ray observation of \swiftOhSevenOne.
The latter four objects are thus confirmed magnetic CVs of the intermediate polar class.
An optical period of 1554~s in \axj\ also confirms the known X-ray spin
period, but a stronger signal at 2303~s is present whose interpretation is
not obvious.  We also studied the candidate intermediate polar \swiftOhEight,
which has low and high states differing by $\approx4$~mag, and
optical periods or QPOs not in agreement with proposed X-ray periods.
Of note is an unusually long 2.06~day orbital period for \swiftOneNineOh,
manifest in the radial velocity variation of photospheric absorption
lines of an early K-type companion star. The star must be somewhat evolved
if it is to fill its Roche lobe.   
\end{abstract}


\section{INTRODUCTION}

This paper is the fourth in a series of optical
studies of cataclysmic variables (CVs) that were discovered
in surveys at hard X-rays energies ($>15$~keV) by the  
\sw\ Burst Alert Telescope (BAT) and the
International Gamma-Ray Astrophysics Laboratory (\int).
CVs are accreting binaries in which a white dwarf (WD)
accretes mass from a low-mass main-sequence or slightly
evolved companion that fills its own Roche-lobe.  If the
magnetic field of the WD is strong enough it will truncate
the inner accretion disk at the WD magnetospheric boundary,
funneling an accretion column directly onto a magnetic pole
or poles.  Even stronger magnetic fields can channel an
accretion stream directly from the surface of the companion,
completely preventing a disk from forming.  Magnetic CVs are
efficient at producing hard X-rays in the accretion column,
where thermal bremsstrahlung X-rays are emitted by
shock-heated plasma just above the surface of the WD.

The systems with partially truncated disks are the 
intermediate polars (IPs, or DQ Herculis stars).
In these, the accretion stream or curtain is channeled from the
inner edge of the disk to the magnetic pole(s).
The spin period of the WD in an IP can be detected
as a coherent X-ray or optical oscillation
arising from the rotating hot spot at the base of the
accretion column, at a shorter period than the orbital
period of the binary.  Sometimes the beat frequency
between the spin and orbit is seen in emission reprocessed
by material in the orbiting frame.  The X-rays often suffer
energy-dependent photoelectric absorption in the accretion curtain.

When the magnetic field is strong enough to completely block the formation of an
accretion disk, it usually also locks the
WD rotation to the binary orbit. These synchronous systems
are the polars (AM Herculis stars).  Polars also exhibit
circularly polarized optical emission and optical/IR
humps in their spectra, both of these being features
of cyclotron radiation in a strong magnetic field. 
There are also a handful of asynchronous polars,
in which the spin and orbit periods differ by up to a few percent.

The properties of hard X-ray selected CVs were most recently
reviewed by \citet{dem20}, \citet{fal19}, and \citet{lut20}.
Among those detected by \sw\ BAT \citep{oh18} and \int\ \citep{bir16},
IPs outnumber all the other subclasses of CVs including polars, 
and continue to dominate as new classifications are made by X-ray and
optical followup studies cited in the above reviews \citep[e.g.,][]{ber17}.
It is thought that IPs outnumber polars in hard X-ray
samples because of their higher accretion rates, and also
because in polars the accretion may be ``blobby,'' depositing
some fraction of its energy directly onto the surface of the
WD, to be radiated at lower temperatures, tens of eV.

We aim to advance the completeness of CV identifications
and classifications from \sw\ BAT and \int.  We measure the
CVs' orbital periods using time-resolved optical
spectroscopy and/or time-series photometry.  Our fast
photometry also reveals spin modulation when present,
allowing for classification of magnetic subclasses of CVs.
In Thorstensen \& Halpern (2013, Paper~I)
Halpern \& Thorstensen (2015, Paper~II), and Halpern \& Thorstensen
(2018, Paper~III), we presented data and findings on 36 hard X-ray
selected CVs, 14 of which were unidentified X-ray sources prior to our work.
The present paper continues this effort with targets selected from prior
X-ray studies, as well as unidentified X-ray sources with counterparts
in pointed observations from the imaging \sw\ X-ray Telescope (XRT).
We present new data on \numberOfTargets\ objects: four newly identified
CV counterparts and six previously studied ones.

\section{EQUIPMENT AND TECHNIQUES}
\label{sec:techniques}

Nearly all our optical data are from the MDM Observatory, which
comprises the 1.3-m McGraw-Hill telescope and the
2.4-m Hiltner telescope, both on the southwest ridge
of Kitt Peak, Arizona.  For the spectroscopy and radial 
velocity studies we used the modular spectrograph (modspec) on both 
the 1.3-m and the 2.4m telescopes, and the Ohio State Multi-Object
Spectrograph (OSMOS; \citealt{osmos}) on the 2.4-m only.
The observing protocols, reduction, and analysis 
techniques we used for the spectroscopy were essentially 
identical to those detailed in Papers I, II, and III.

High-cadence differential photometry sensitive to spin periods
was carried out with the MDM 1.3-m, or, for one target,
at the South African Astronomical Observatory (SAAO) 
1-m telescope, using the Sutherland High-speed
Optical Camera \citep{shoc}. Again, our techniques are 
detailed in earlier papers.

\section{RESULTS ON INDIVIDUAL OBJECTS}
\label{sec:individuals}

Table~\ref{tab:objects} lists the objects studied here,
with accurate celestial coordinates,
approximate magnitudes from Pan-STARRS, and periods where determined.  A photometric period is considered confirmed only if it is detected independently on different nights.  Uncertainty in period $P$ is determined from the width of the periodogram peak at power level $S_{\rm max}-1$, where $S_{\max}$ is the peak power value.  In practice, this corresponds to $\sigma(P)\approx0.1\,P^2/T$, where $T$ is the time span of the data from beginning to end, and the numerical factor is accurate to $\sim50\%$.

\begin{figure}
\vspace{-1.0in}
\centerline{
\includegraphics[angle=0,width=1.\linewidth]{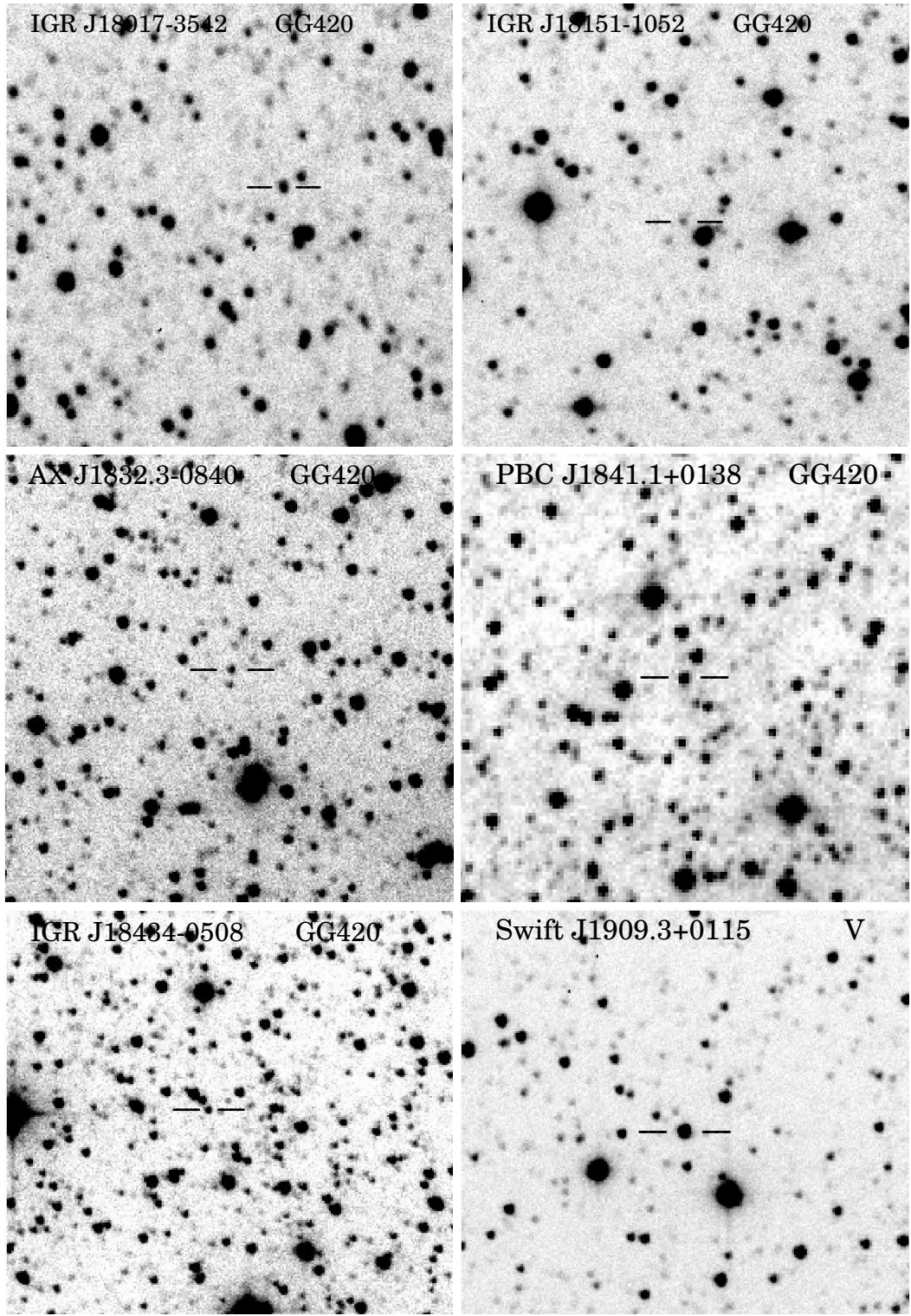}
}
\caption{Finding charts for newly studied objects in Table~\ref{tab:objects},
taken with the MDM 1.3-m; each field is $2.\!^{\prime}2\times2.\!^{\prime}2$.
North is up and east is to the left.}
\label{fig:charts4}
\end{figure}

\begin{figure}
\centerline{
\includegraphics[angle=0,width=1.\linewidth]{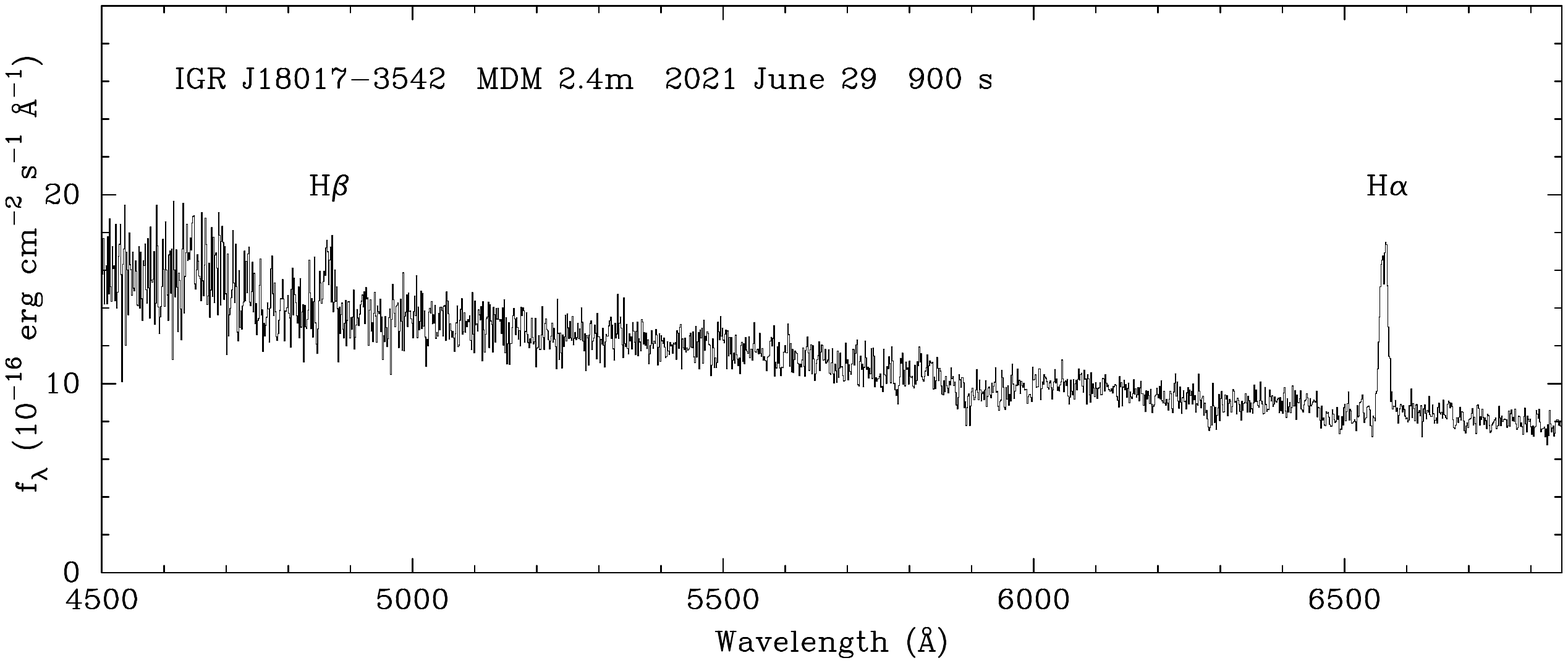}
}
\vspace{-2.7in}
\centerline{
\includegraphics[angle=0,width=1.\linewidth]{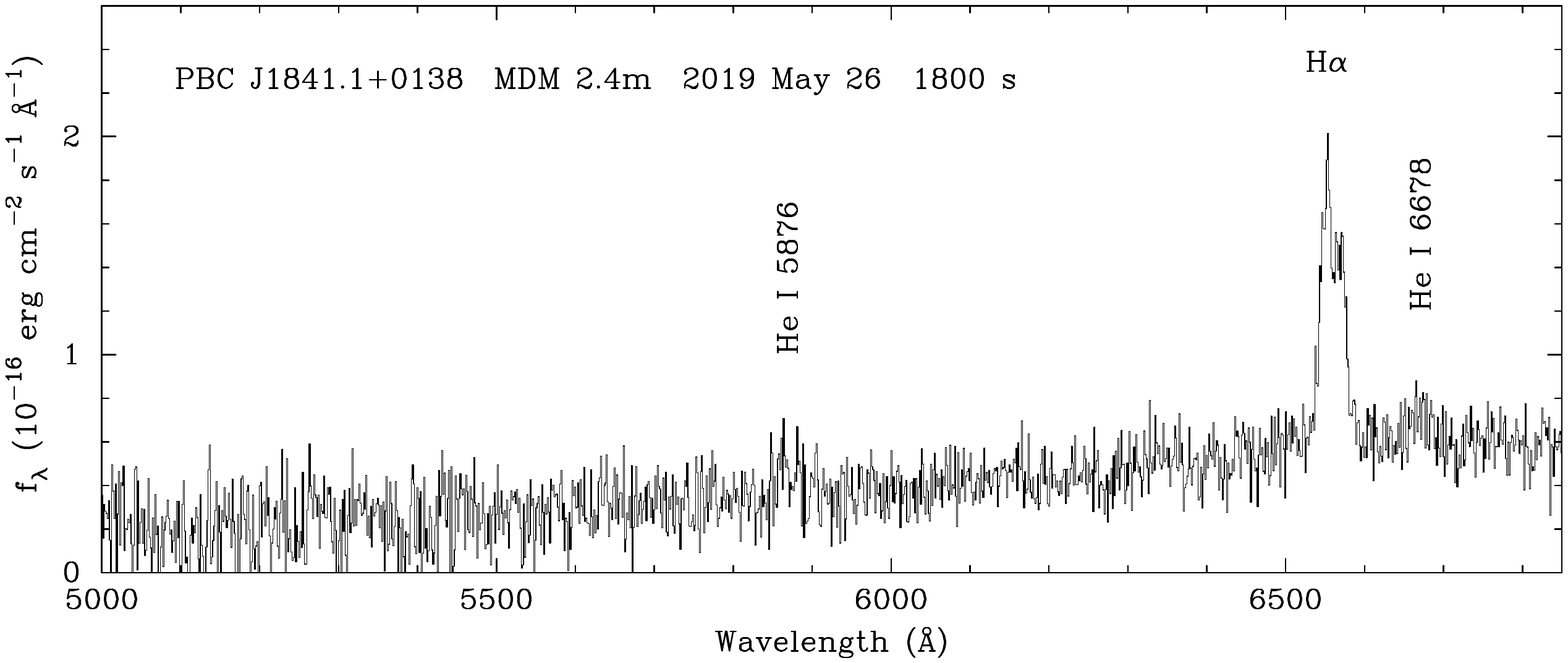}
}
\vspace{-2.7in}
\centerline{
\includegraphics[angle=0,width=1.\linewidth]{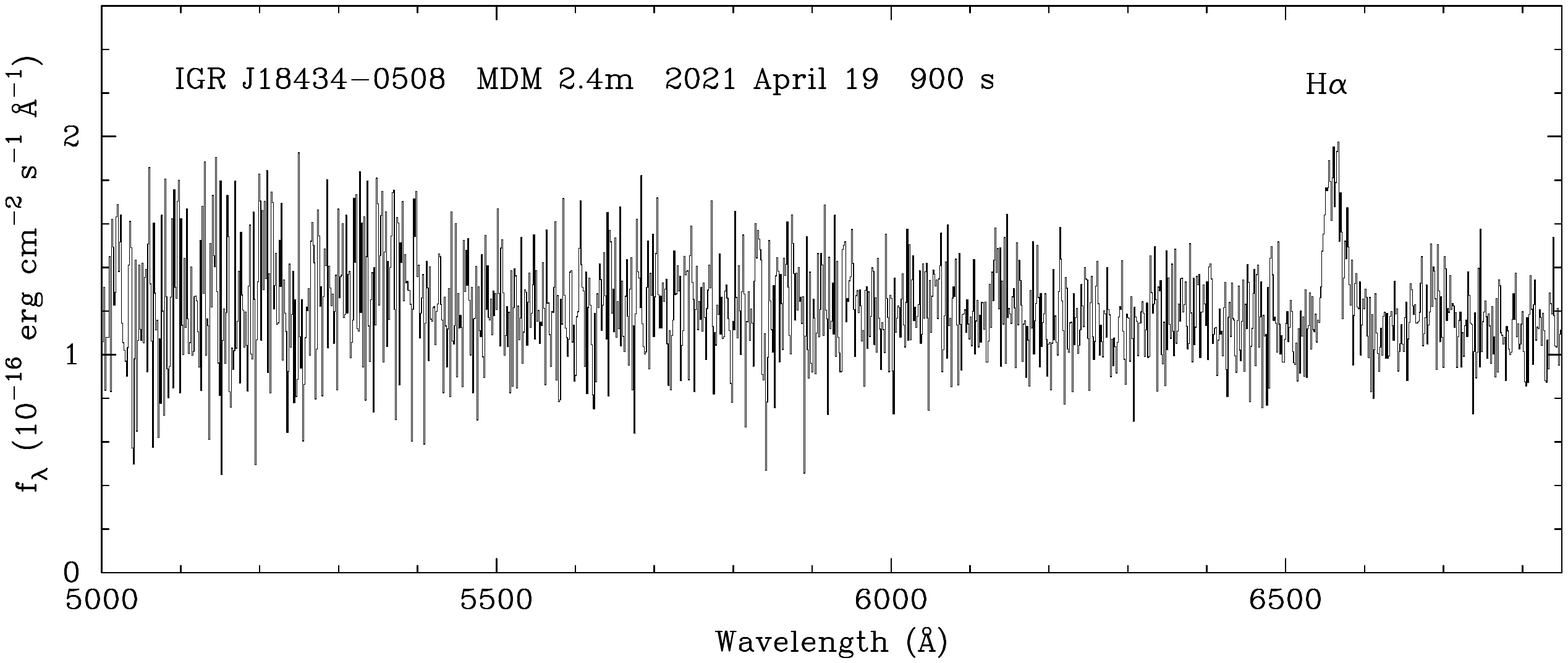}
}
\vspace{-2.6in}
\caption{Spectroscopic identifications of \igrOneEightOh, \pbcOneEight,
and \igrOneEightFour\ using OSMOS.
}
\label{fig:spectrum}
\end{figure}

Figure~\ref{fig:charts_old} shows finding charts for three objects
that were discussed in our previous papers and have new data
reported here.  Figure~\ref{fig:charts4} presents finding charts for
six newly studied (by us) optical counterparts.
Figure~\ref{fig:spectrum} shows the spectra of three of the newly identified sources.  Subsequent Tables and
Figures contain the results and analysis of photometry and
spectroscopy of individual objects.

\subsection{\swiftOhSevenOne\ and \swiftOhSevenFour}
\label{sec:swiftOhSeven}

\begin{figure}
\centerline{
\includegraphics[width=1.1\linewidth]{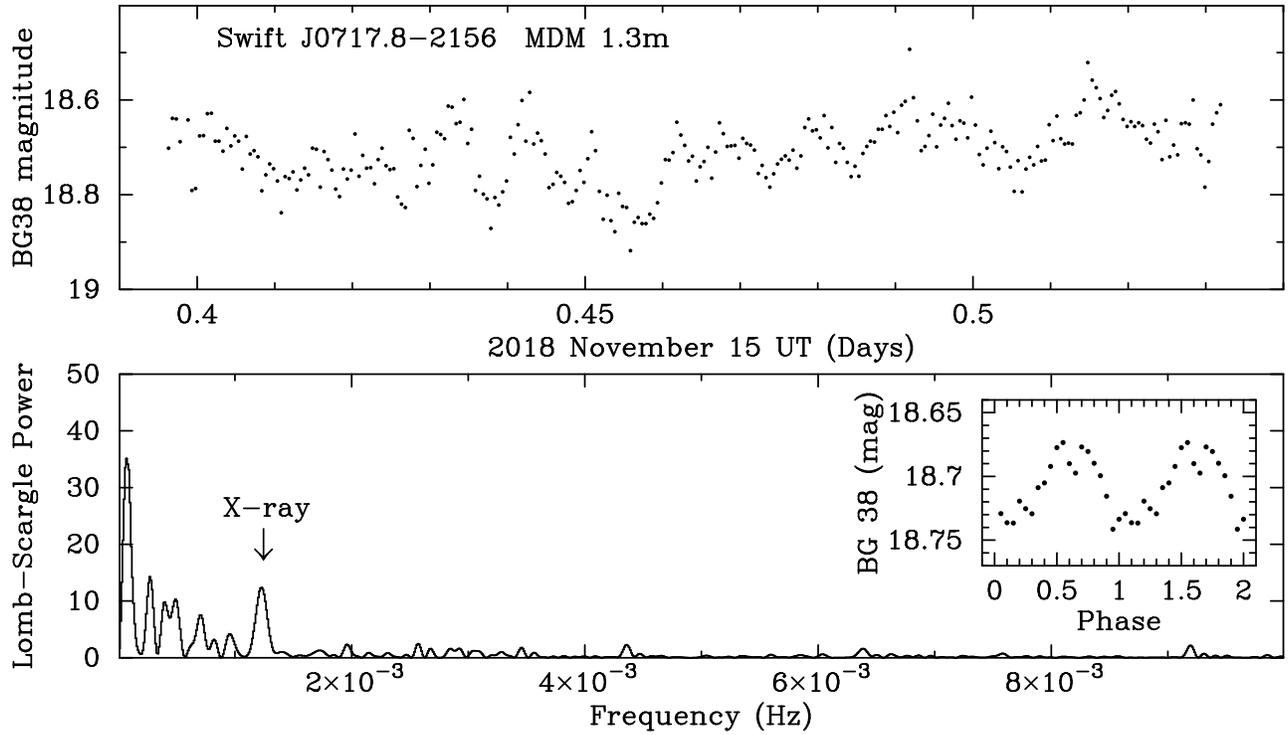}
}
\vspace{-6.in}   
\caption{
Time-series photometry of \swiftOhSevenOne.
Individual exposures are 40~s.  The 803.5~s period found in \xmm\ data (Figure~\ref{fig:xmm})
is indicated by an arrow, and the optical data are folded at that period (inset).
}
\label{fig:swiftj0717}
\end{figure}

\begin{figure}
\centerline{
\includegraphics[width=1.1\linewidth]{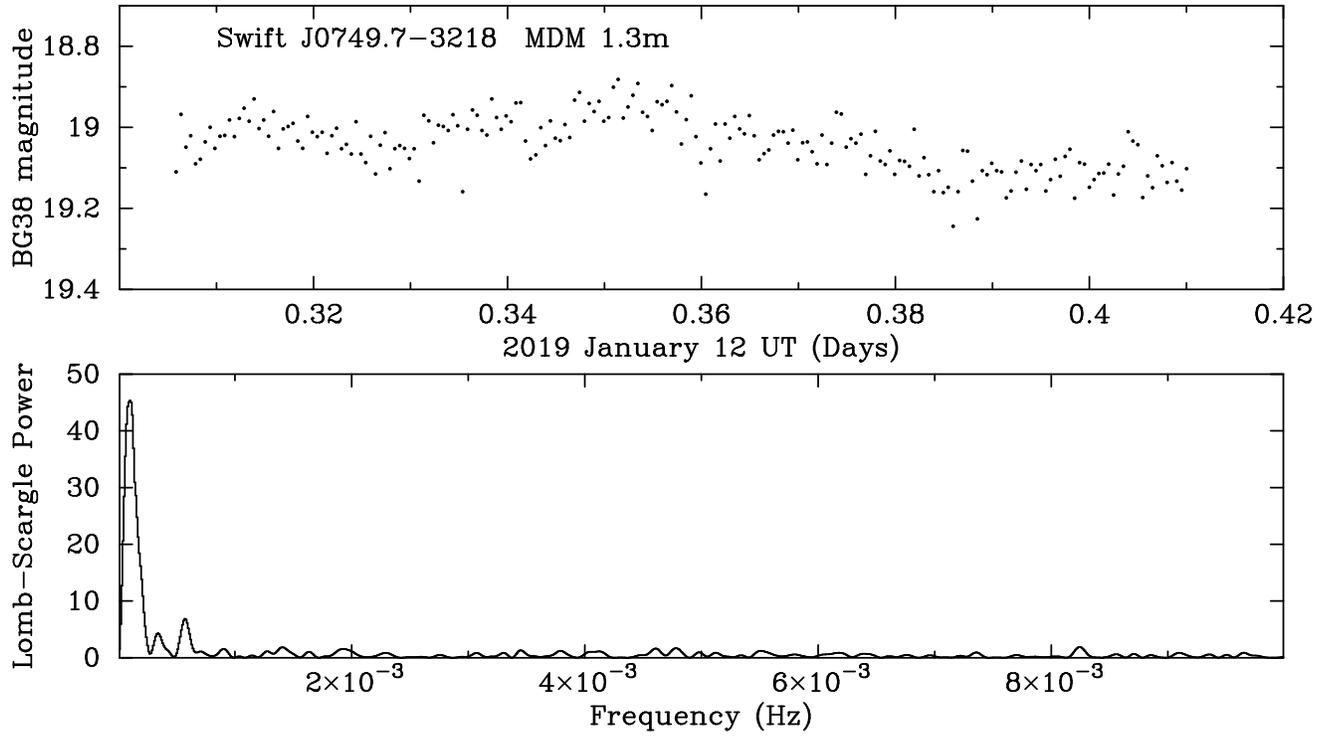}
}
\vspace{-6.in}   
\caption{
Time-series photometry of \swiftOhSevenFour.
Individual exposures are 40~s.
}
\label{fig:swiftj0749}
\end{figure}

\swiftOhSevenOne\ and \swiftOhSevenFour\ were identified spectroscopically
as likely CVs in Paper~II, based on their Balmer and \ion{He}{1} emission
lines.  But no time-series data had been obtained at that time.
In Paper~III, a radial velocity study of \swiftOhSevenOne\ established an
orbital period of $0.2298\pm0.0010$~days, and detected
\ion{He}{2}~$\lambda4686$, suggesting that it is a magnetic system.
We subsequently obtained a short time-series for each of these stars
that confirms their CV nature through flickering, but does not detect
any significant periodicity (Figures~\ref{fig:swiftj0717} and \ref{fig:swiftj0749}).

\begin{figure}
\vspace{-1.in}
\centerline{
\includegraphics[width=1.4\linewidth]{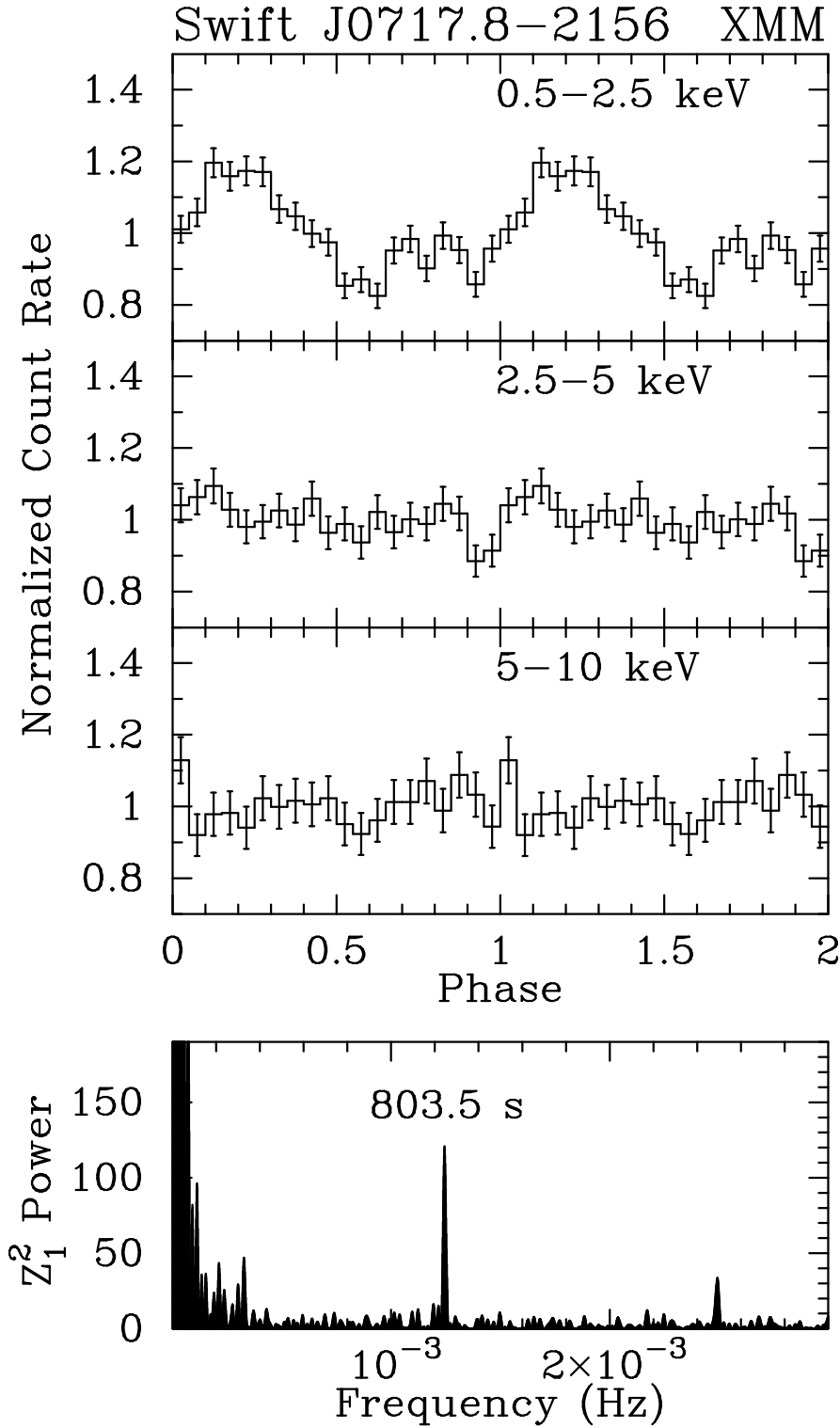}
}
\vspace{-3.2in}
\caption{
Top: Energy-dependent folded light curves from the \xmm\ observation of
\swiftOhSevenOne\ on 2019 March~28.  Background from a nearby region
in the image has been subtracted, and the counts per bin are 
normalized to 1.  Bottom: Power spectrum of the 0.5--2.5~keV
counts showing a period at $803.5\pm0.7$~s, and its first harmonic.
}
\label{fig:xmm}
\end{figure}

However, we report here the discovery of a likely spin period in an
archival X-ray observation of \swiftOhSevenOne.  Taken on 2019 March~28
by \xmm\ (ObsID~0820330801), it has a continuous exposure of 49~ks.
Figure~\ref{fig:xmm} shows the folded light curve from the combined 
pn and MOS photon lists in three energy bands, and a $Z_1^2$ periodogram
of the lowest (0.5--2.5~keV) energy band, where the modulation is prominent.
A period of $803.5\pm0.7$~s is detected, as well as significant
power in the first harmonic due to the asymmetric, double-peaked pulse.
Confined to energies below 2.5~keV,
the pulsation is
likely due to spin-modulated photoelectric absorption
by the accretion curtain, a common effect in IPs.  Examining the
power spectrum of the optical light curve (Figure~\ref{fig:swiftj0717}),
there is a marginally significant peak at the X-ray period.

\subsection{\swiftOhEight}
\label{sec:swiftOhEight}

\begin{figure}
\centerline{
\includegraphics[width=1.1\linewidth]{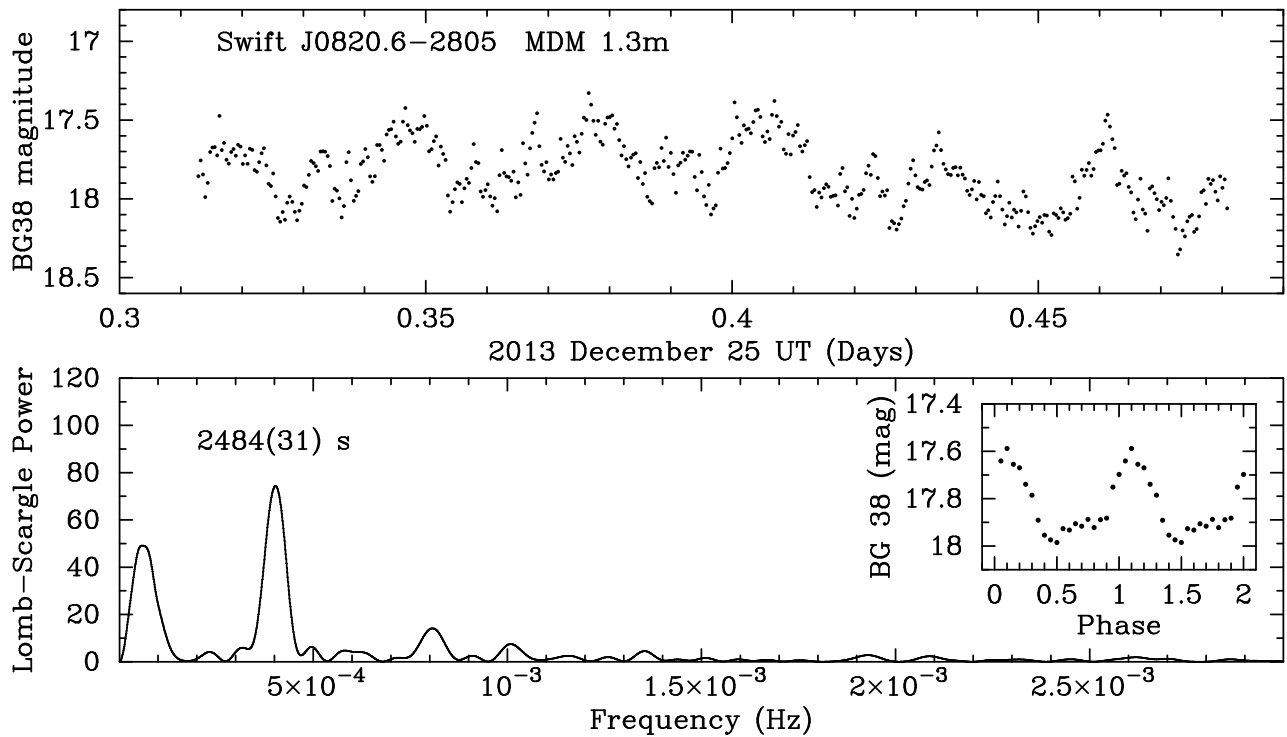}
}
\vspace{-6.in}   
\caption{
Time-series photometry of \swiftOhEight\ from 2013,
as appeared in Paper II.  Individual exposures are 30~s.
The Lomb-Scargle periodogram identifies a period at $2484\pm31$~s.
}
\label{fig:old}
\end{figure}

\swiftOhEight\ was identified in Paper II, where a candidate period
of $2484\pm31$~s was noted based on a single time series in 2013 December.
Figure~\ref{fig:old} is a reproduction of that analysis.  It was noted
that the period should be considered tentative because the light
curve covers only five cycles of the oscillation. In subsequent
years we attempted to test this result, only to discover that the star
had had faded by several magnitudes and was undetectable in short
exposures.  

\begin{figure}
\vspace{-1.0in}
\centerline{
\includegraphics[angle=0,width=1.\linewidth]{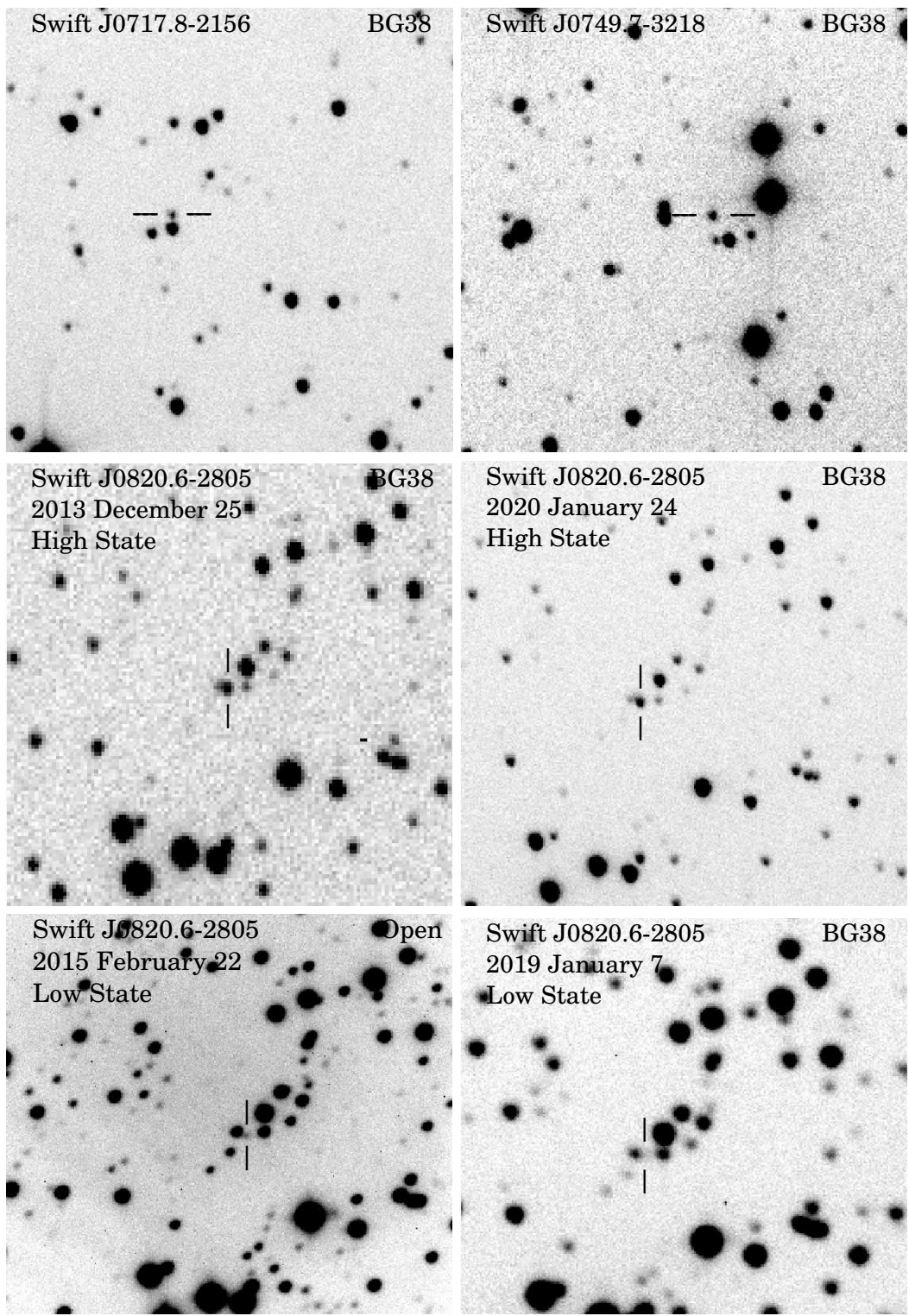}
}
\caption{Finding charts for several of the objects in Table~\ref{tab:objects}
that were studied in our previous papers, but with new data reported here.
All but one were taken with the MDM 1.3-m; the 2015 February image of
\swiftOhEight\ is from the SAAO 1-m.
Each field is $2.\!^{\prime}2\times2.\!^{\prime}2$.
North is up and east is to the left.}
\label{fig:charts_old}
\end{figure}

\begin{figure}
\centerline{
\includegraphics[width=1.1\linewidth]{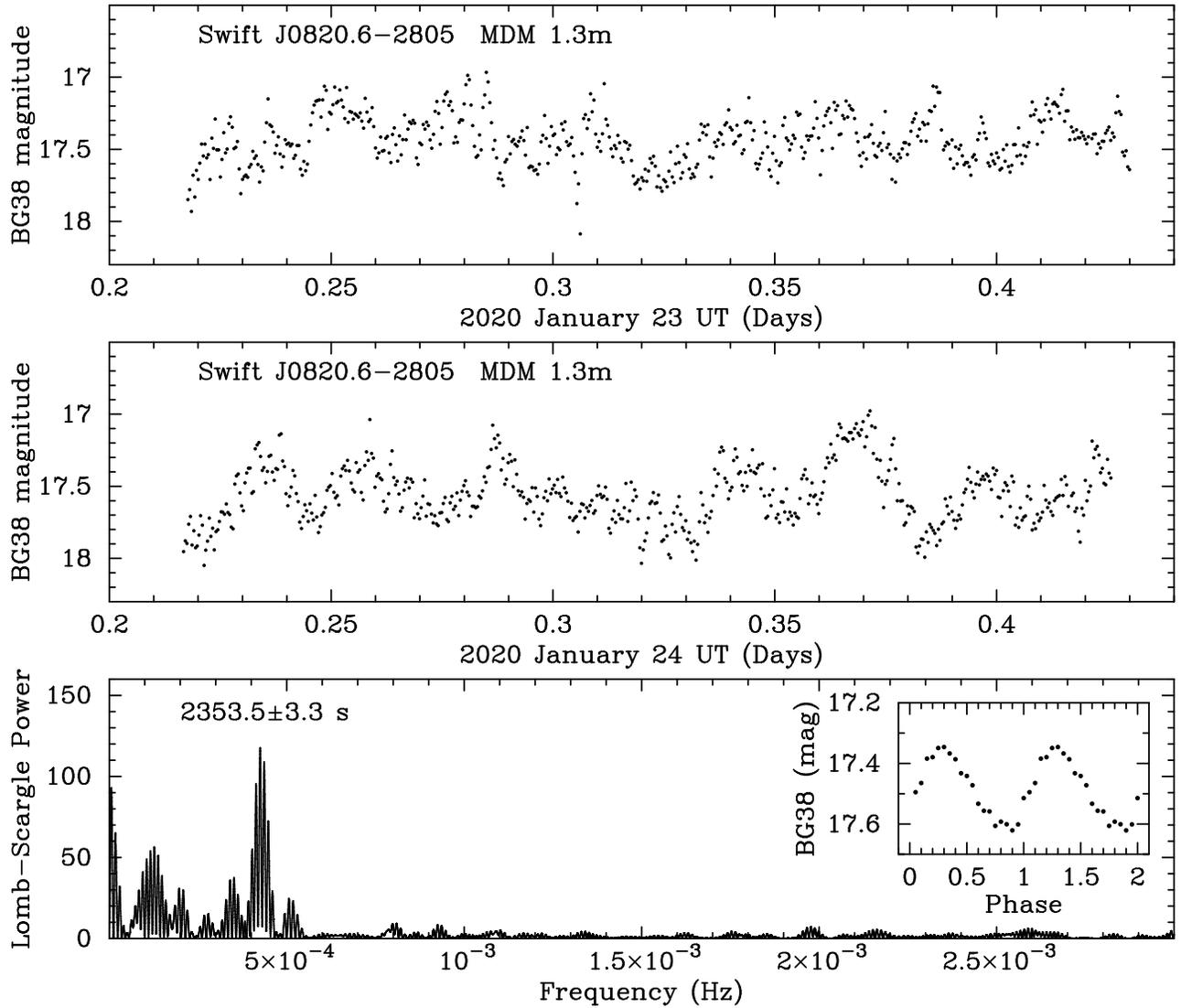}
}
\vspace{-4.in}   
\caption{
Time-series photometry of \swiftOhEight\ from 2020 January 23 and 24.
Individual exposures are 30~s. The joint Lomb-Scargle periodogram
identifies a period at $2353.5\pm3.3$~s, close to but not consistent
with the $2484\pm31$~s period in 2013 (Figure~\ref{fig:old}).
}
\label{fig:swift_joint}
\end{figure}

Figure~\ref{fig:charts_old} shows deeper images taken on
2015 February 22 and 2019 January 7, with the SAAO 1-m 
and the MDM 1.3-m, respectively.  The former, in particular, is
the sum of 1920~s of unfiltered exposures in which the optical counterpart
of \swiftOhEight\ is faintly visible (and is easily confused with a neighboring
star).  Comparing with the $r$ magnitudes of nearby stars in Pan-STARRS,
we conclude that \swiftOhEight\ had faded to $r\approx21$.  (The Pan-STARRS
mean magnitude of \swiftOhEight\ is $r=17.07$.)  Finally,
in 2020 January we caught the star back in a high state at magnitude
$\approx 17.5$ in a BG38 filter, as shown in Figure~\ref{fig:charts_old} (image) and
Figure~\ref{fig:swift_joint} (time series).

With two nights in this high state, the star again shows oscillations,
but the best fitted period is now $2353.5\pm3.3$~s, which is close to
but not consistent with the original value of $2484\pm31$~s.  Therefore,
we cannot consider this confirmation.  It could be that both of these
signals are simply transient quasi-periodic oscillations.  Alternatively,
if one signal represents the spin period, and the other a beat period
between the spin and the orbit, it suggests an orbital period of
$\approx0.52$~days.  This is not an implausible value, but as an explanation
this requires a change of accretion geometry such that the direct emission
from the spinning WD is stronger at one epoch, while reprocessed light
from an orbiting structure dominates at the other.  Further observations
would be needed to investigate the origin of these signals.

\citet{nuc20} analyzed a 37~ks \xmm\ observation of \swiftOhEight\ that was
taken in 2018.  It was clearly in a low state then, since its average
X-ray count rate was only 0.015~s$^{-1}$, corresponding to a 0.3--10~keV
flux of $\sim10^{-13}$ erg~cm$^{-2}$~s$^{-1}$. This is to be compared with
the count rate of $0.2-0.3$~s$^{-1}$ in the much less sensitive Swift XRT
in 2009 and 2011, corresponding to $\sim10^{-11}$ erg~cm$^{-2}$~s$^{-1}$.
In addition its average $B$ magnitude in the \xmm\ optical monitor was 20.49,
close to the minimum we observed in 2015 and 2019.  Nevertheless,
\citet{nuc20} identified peaks in the X-ray periodogram
at 27.87~min and 87.53~min as the spin and orbital periods.  Neither of
these correspond to our 41.4~min (2484~s) and 39.2~min (2353~s) periods. 
The peaks in the X-ray periodogram are weak; we suggest that a similar X-ray 
observation performed when the source is in its high state will detect any
true period(s) and address the present ambiguities.

\subsection{\igrOneEightOh}
\label{sec:igrOneEightOh}

This \int\ source from \cite{kri17} was observed with \chandra\ by \cite{tom21},
who identified its probable X-ray and optical counterpart.  Based on its optical/IR
magnitudes, Gaia parallax and proper motion, and X-ray luminosity, \cite{tom21} classified it as
a likely CV.  We obtained a spectrum of this star with OSMOS on 2021 June 26 that
confirms it as a CV based on its broad, double-peaked H$\alpha$ emission line 
and blue continuum (Figure~\ref{fig:spectrum}).

\begin{figure}
\centerline{
\includegraphics[width=1.1\linewidth]{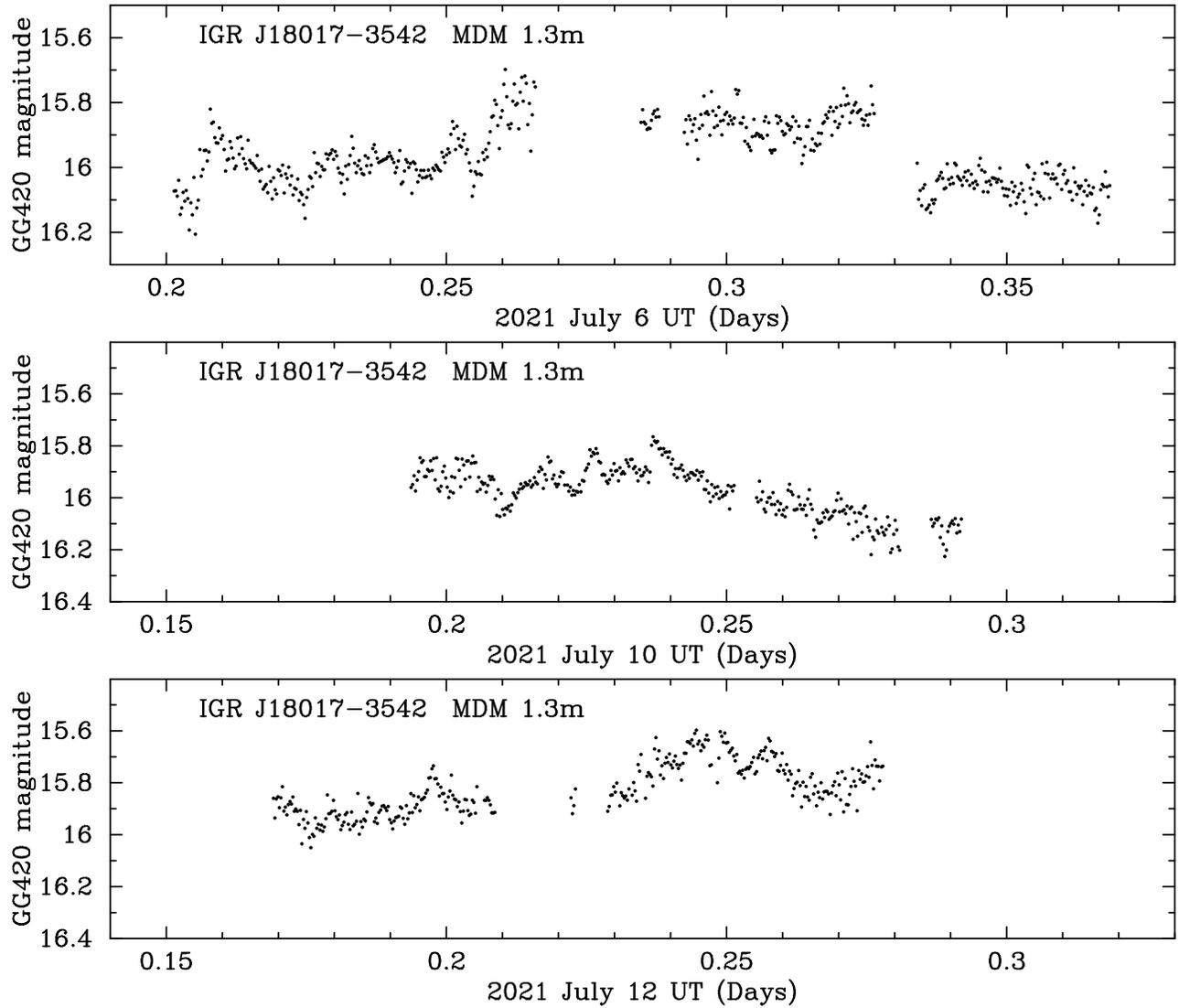}
}
\vspace{-4.in}   
\caption{
Time-series photometry of \igrOneEightOh\ in 2021 July
Individual exposures are 20~s. Frequent gaps are due to clouds.
}
\label{fig:igrOneEightOh}
\end{figure}

We then followed up with time-series photometry through a GG420 filter on three nights in
2021 July (Figure~\ref{fig:igrOneEightOh}).  Although displaying flickering up to 0.5~mag,
no periodicity is detected in the power spectra (not shown).  However, the time-series
were frequently interrupted by clouds, and longer contiguous runs,
especially from the southern hemisphere, would be of interest for this
bright, $\approx16$~mag star.

\subsection{\igr}
\label{sec:igr}

The optical identification of \igr\ was made by \citet{lut12a} with
the aid of a Swift XRT position.  Their optical spectrum has a strong H$\alpha$
emission line which, together with a hard X-ray spectrum, favored classification
as a CV. \citet{mas13} also obtained an optical spectrum, which
shows, in addition to H$\alpha$, emission lines of H$\beta$,
\ion{He}{2}~$\lambda4686$ and the Bowen \ion{C}{3}/\ion{N}{3} blend, 
which further suggested a magnetic CV.  Recently, \citet{wor20} obtained
an \xmm\ observation that clearly showed a 390~s period, interpreted as
the WD spin.

\begin{figure}
\centerline{
\includegraphics[width=1.1\linewidth]{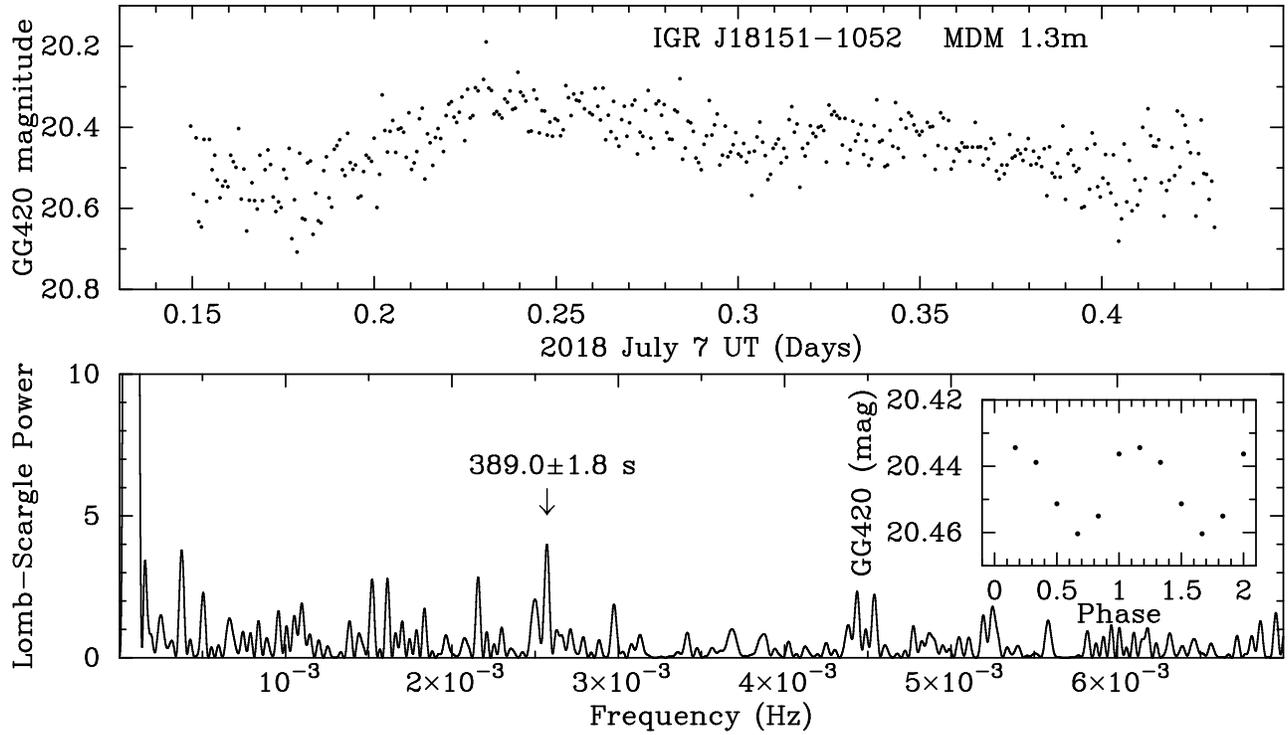}
}
\vspace{-6.in}   
\caption{
Time-series photometry of \igr.  Individual exposures are 60~s.
The Lomb-Scargle periodogram identifies a weak signal at 389~s,
consistent with the X-ray period of \citet{wor20}.
}
\label{fig:igrj18151}
\end{figure}

We obtained a single time-series observation of \igr\ through a GG420 filter
with a cadence of 63~s (Figure~\ref{fig:igrj18151}).  Although not providing
a significant independent detection of the 390~s period, we note that the
highest peak in the Lomb-Scargle periodogram occurs at $389.0\pm1.8$~s,
consistent with the X-ray period.
A light curve folded at this period has an amplitude of $\approx0.03$~mag.

\subsection{\axj} 
\label{sec:axj}

\axj\ was discovered as a pulsating X-ray source by \citet{sug00}.
Its period of $1549.1\pm0.4$~s together with a 6.7~keV iron line and
luminosity based on an estimated distance of $\le4$~kpc
identified it as a likely magnetic CV.
It is also detected by \int\ \citep{bir16} and by Swift-BAT \citep{oh18}
as Swift J1832.5$-$0863 (a non-standard coordinate naming convention).
The X-ray properties of \axj\ were studied in more detail by
\cite{kau10} using \xmm. Its features
include a three-component Fe line complex, supporting classification
as an IP together with their measured spin period of $1552^{+2.3}_{-0.8}$~s.
\cite{kau10} also used \chandra\ to get a precise position,
which coincided with an optical/IR counterpart
whose basic properties are listed in Table~\ref{tab:objects}.
\citet{mas13} obtained an optical spectrum showing Balmer emission lines
and \ion{He}{2}~$\lambda$4686, confirming \axj\ as a magnetic CV.

\begin{figure}
\centerline{
\includegraphics[width=1.1\linewidth]{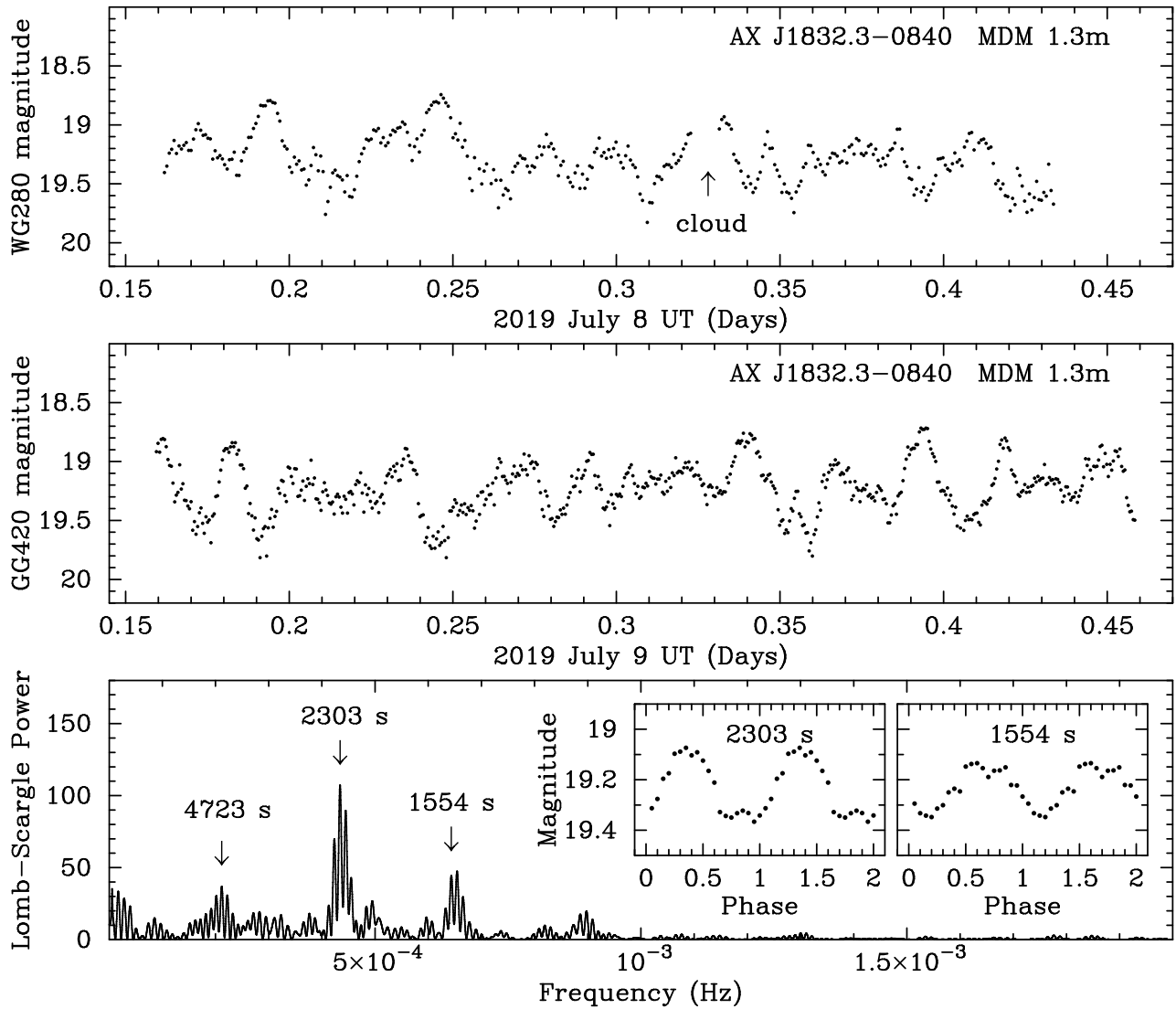}
}
\vspace{-4.in}   
\caption{
Time-series photometry of \axj.  Individual exposures
are 60~s on July~8, and 40~s on July~9.  The Lomb-Scargle periodogram
identifies a period at 1554~s, consistent with the X-ray period,
and a signal with higher power at 2303~s.
Light curves folded on both periods are shown.
}
\label{fig:axj1832}
\end{figure}

We obtained time-series photometry on two adjacent nights, 2019 July 8 and 9,
to further study periodic phenomena in \axj.  A clear WG280 filter
was used on the first night, with exposures of 60~s.  On the second night
a GG420 filter was used to mitigate against moonlight, and the
exposure time was reduced to 40~s.  Figure~\ref{fig:axj1832} shows the
resulting light curves and Lomb-Scargle periodogram.  Interestingly,
while there is a peak at 1554~s (and its 1-day alias at 1529~s) matching
the X-ray period, a stronger signal is seen at a period of 2303~s,
and possibly a weaker one at 4723~s.  None of these are exact multiples
of each other, except for an alias peak at 4500~s which is close to
twice one at 2248~s.  

While it is very likely that the X-ray period represents
the spin, the interpretation of the other optical periods
is not clear, and we do not have a spectroscopic orbital period.

\subsection{\pbcOneEight}
\label{sec:pbc}

The Swift XRT targeted a source named \pbcOneEight\ twice in 2019 February.
Although the Palermo Swift-BAT Catalog of hard X-ray sources has
been updated several times, we can find no record of a source with this name in any publication or public database.
However, an unidentified BAT source, Swift J1841.0+0152, is
listed with coordinates (J2000) R.A.=$18^{\rm h}41^{\rm m}08.\!^{\rm s}7$,
decl.=$+01^{\circ}39^{\prime}57^{\prime\prime}$ in \citet{oh18}, which is
within $2^{\prime}$ of the coordinate designation of \pbcOneEight.
Therefore, we assume that these BAT sources are one and the same.

In a total exposure of 2437~s, the XRT detected a point source
of 37 relatively hard photons at position consistent with \pbcOneEight,
and within $2^{\prime\prime}$ of the faint optical object identified in
Figure~\ref{fig:charts4}, whose basic properties
are listed in Table~\ref{tab:objects}.  A spectrum of this star
obtained with OSMOS on 2019 May 26 identifies it as a probable CV based on its broad, double-peaked H$\alpha$ emission and weaker \ion{He}{1} emission lines (Figure~\ref{fig:spectrum}).

\begin{figure}
\vspace{-0.4in}
\centerline{
\includegraphics[width=1.1\linewidth]{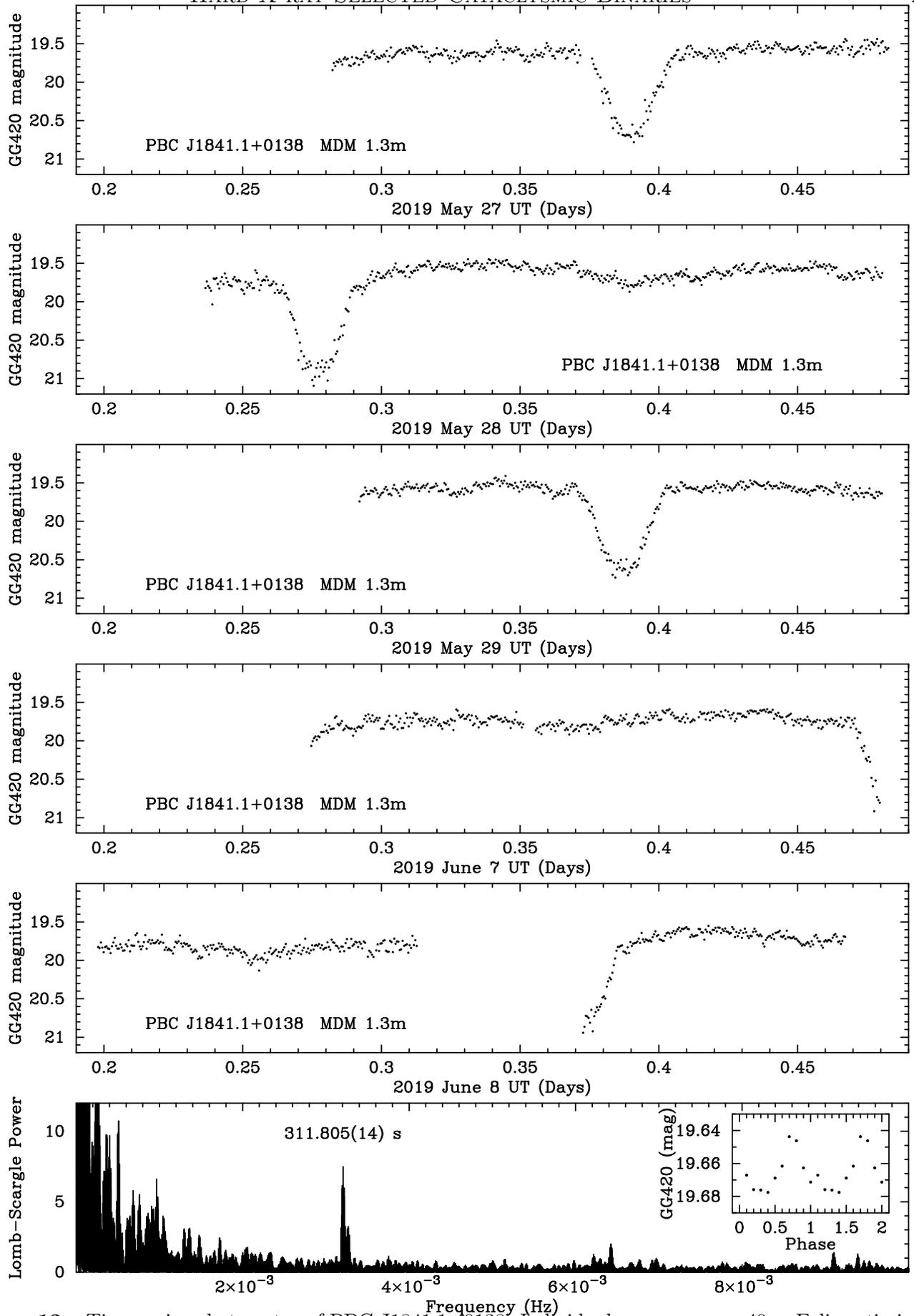}
}
\vspace{-0.4in}   
\caption{
Time-series photometry of \pbcOneEight.  Individual exposures
are 40~s. Eclipse timings are listed in Table~\ref{tab:eclipse}.
Eclipses were excised before calculating the Lomb-Scargle periodogram.
A signal at 311.8~s is detected, as well as its first harmonic.
}
\label{fig:pbcj1841}
\end{figure}

We then followed up with time-series photometry of \pbcOneEight\
through a GG420 filter on five nights in
2019 May and June (Figure~\ref{fig:pbcj1841}).
On each night there is a clear eclipse,
deeper than 1 magnitude, although on two of the nights only ingress
or egress was covered. An unambiguous eclipse frequency of 4.5 cycles day$^{-1}$
is evident, which made it possible to schedule two brief time-series
(not shown) covering eclipses on June 28 and 29 to further 
refine the orbital ephemeris.

From the five complete eclipses observed,
times of mid-eclipse were derived by fitting a second-order
polynomial to the magnitudes in each dip.  The mid-eclipse
times (listed in Table~\ref{tab:eclipse}) were then fitted to a constant
orbital period leaving timing residuals $<15$~s, smaller
than the 43~s cadence of the data.
The eclipse ephemeris (in TDB) spanning JD 2458630--2458663 is
$$T_{\rm mid} =  {\rm BJD}\ 2458630.88977(10) + 0.2219088(11)\,E.$$
In addition to the eclipses, the light curves have much shallower,
broad modulation that could be due to some combination of ellipsoidal
modulation and partial eclipse of the secondary star by the accretion disk.

\begin{figure}
\centerline{
\includegraphics[width=1.1\linewidth]{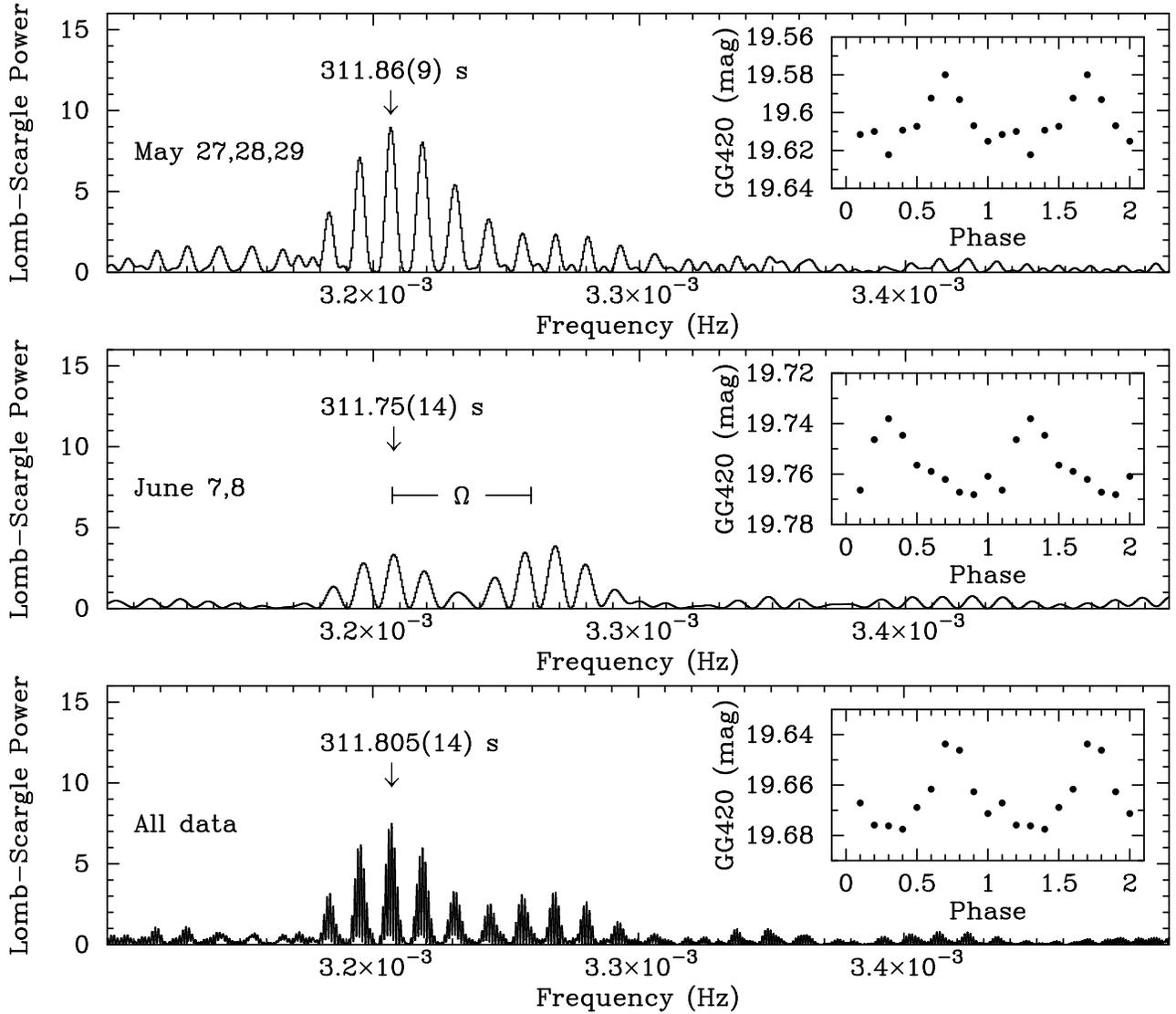}
}
\vspace{-3.9in}   
\caption{
Expanded periodograms of the light-curves of \pbcOneEight\ shown
in Figure~\ref{fig:pbcj1841}.  The data have been grouped into
adjacent nights in May (top) and June (middle).  The bottom
panel is the joint periodogram of all five nights 
as in Figure~\ref{fig:pbcj1841}.
In each panel the best fitted period of the highest peak is
indicated.  The length of horizontal bar in the middle panel
represents the orbital frequency $\Omega$, which could be the
difference between the spin frequency $\omega$ 
and the spin-orbit beat frequency $\omega-\Omega$.
}
\label{fig:power}
\end{figure}

The power spectra also show a persistent,
coherent period of 311.8~s with full amplitude of $\approx0.04$~mag,
likely related to the spin of the WD.
In order to measure the period most precisely, we excised the eclipses
from the light curves and calculated a Lomb-Scargle
periodogram of the remaining time series, which is shown in the
bottom panel of Figure~\ref{fig:pbcj1841}.
Expanded versions of the periodogram around the resulting $311.805\pm0.014$~s
signal are shown in Figure~\ref{fig:power}, where the data sets are grouped into
adjacent nights (May and June separately), and the full set of five.

These periodograms show a persistent main peak at 311.8~s, with weaker
1-day aliases separated by $\approx1$~s.
But there is also structure at slightly higher frequencies,
which suggests that another period is present.  In the
June data, a secondary peak at either 306~s or 307~s is actually
stronger than the main peak. This is reminiscent of many intermediate
polars in which both the spin frequency $\omega$ and the difference,
or beat between the spin and orbit, $\omega-\Omega$, can be present
in the optical power spectrum to varying degrees due to reprocessing
of the beamed emission from the WD on material fixed in the orbiting
frame.

This explanation requires that orbital frequency $\Omega$ is the
difference between closely spaced peaks,
so we represent this difference with a horizontal bar marked $\Omega$
in the middle panel of Figure~\ref{fig:power}.
The beat-frequency hypothesis can account approximately
for the dual structure in the periodogram, but perhaps not exactly,
as the peaks don't precisely align when assuming the precisely
known $\Omega$ as the difference. In addition, it is not clear
which peak represents the spin frequency.  Normally the spin frequency
is higher than the reprocessed one due to prograde spin and orbit,
but sometimes $\omega$ and $\omega + \Omega$ can occur.
For definiteness, we list in Table~\ref{tab:objects} 
the $311.805\pm0.014$~s period of the peak that is 
unambiguously strongest in the merged periodogram, though
it is not certain whether this is the spin or the beat period.
Finally, we note that a 10-day aliased value of 311.912~s
is also possible for this peak.

\subsection{\igrOneEightFour}
\label{sec:igrOneEightFour}

\igrOneEightFour\ from \cite{kri17} was observed with \chandra\ by \cite{tom21},
who identified its probable X-ray and optical counterpart and classified it as
a likely CV.  We obtained a spectrum of this star with OSMOS on 2021 April 19 that
confirms it as a CV based on its broad H$\alpha$ emission line (Figure~\ref{fig:spectrum}).

\begin{figure}
\centerline{
\includegraphics[width=1.1\linewidth]{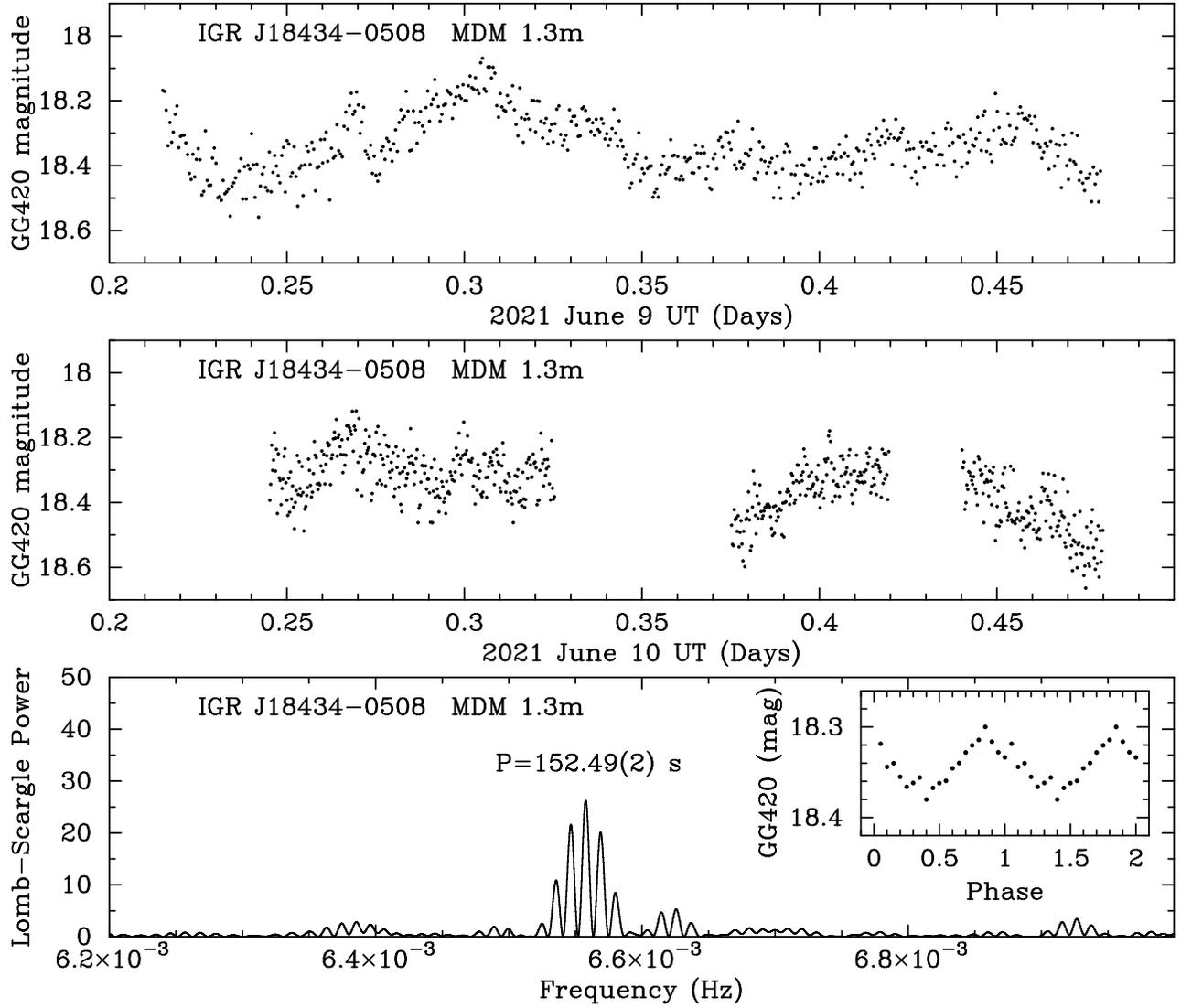}
}
\vspace{-4.in}   
\caption{
Time-series photometry of \igrOneEightFour\ on two consecutive nights
in 2021.  Individual exposures are 40~s on June 9 and 20~s on June 10.
The joint Lomb-Scargle periodogram identifies a period at $152.49\pm0.02$~s.
}
\label{fig:igrOneEightFour}
\end{figure}

On the nights of 2021 June 9 and 10 UT we obtained time-series photometry
of \igrOneEightFour\ through a GG420 filter (Figure~\ref{fig:igrOneEightFour}).
The star shows flickering with an amplitude of $\approx0.3$ mag.
On each night we also detect coherent pulsations with a period of 152.5~s
and full amplitude of $\approx0.08$~mag, likely the spin signature of an IP.
Given the short period, we reduced the individual exposure times from 40~s
on June 9 to 20~s on June 10 in order to better resolve the pulse.  But
the star is too faint to see individual pulses, and the folded pulse profiles
look similar on both nights.
A coherent analysis combining both nights' data derives a period of $152.49\pm0.02$~s.
No other periods are detected in the full periodogram.

\subsection{\swiftOneNineOh}
\label{sec:swift}

\swiftOneNineOh\ is found in the 
Swift-BAT 105-month survey \citep{oh18}.
It was also targeted six times by the Swift XRT between
2017 November 24 and 2018 March 7, under the coordinate
name Swift J1909.3+0115. (\citealt{oh18},
on the other hand, used a nonstandard coordinate naming convention.)
In each XRT observation, a variable source was detected coincident
with the bright optical counterpart listed in Table~1.
Its maximum 0.3--10 keV count rate is $\sim0.15$~s$^{-1}$, with
an average of about half that.  Several of these images are heavily
contaminated by scattered X-rays from the bright source Aql X-1, which is
$23^{\prime}$ away.  The Swift-BAT source was also identified with
XMMSL2 J190921.2+011225 and was suggested to be Galactic by
\citet{ste18}.  ROSAT X-ray designations
are 1RXS J190923.2+011154 and 1WGA J1909.3+0112; evidently
these are all the same object.
We reported briefly on our optical identification of \swiftOneNineOh\ in \citet{hal18}, and present more complete results here.

Table \ref{tab:swiftobsjournal} summarizes our spectroscopic observations.
A single discovery spectrum obtained 2018 March 4 showed
bright emission lines of H, \ion{He}{1}, \ion{He}{2}, and the
\ion{C}{3}/\ion{N}{3} Bowen blend typical of a cataclysmic variable,
together with (more rarely seen) photospheric absorption features of a K-type companion star.

\begin{figure}
\vspace{-0.2in}
\hspace{-0.4in}
\centerline{
\includegraphics[width=1.2\linewidth]{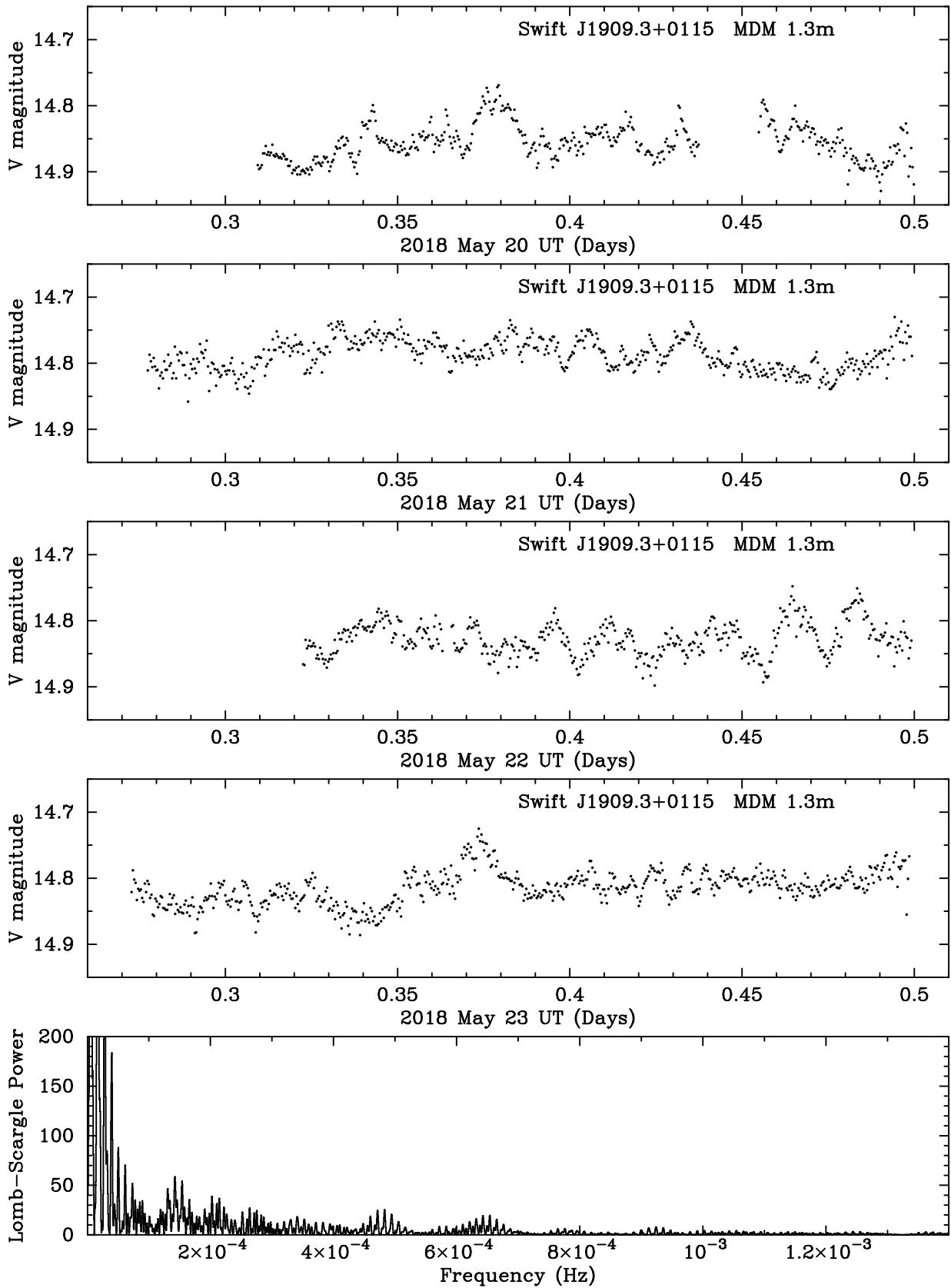}
}
\vspace{-1.9in}   
\caption{
Time-series photometry and Lomb-Scargle periodgram of \swiftOneNineOh\
from four consecutive nights in 2018 May.  Individual exposures are 30~s.
The joint Lomb-Scargle periodogram shows flickering on a range of timescales, 
but does not reveal a clear, coherent period.}
\label{fig:swiftj1909}
\end{figure}

\begin{figure}
\hspace{-0.4in}
\centerline{
\includegraphics[width=1.2\linewidth]{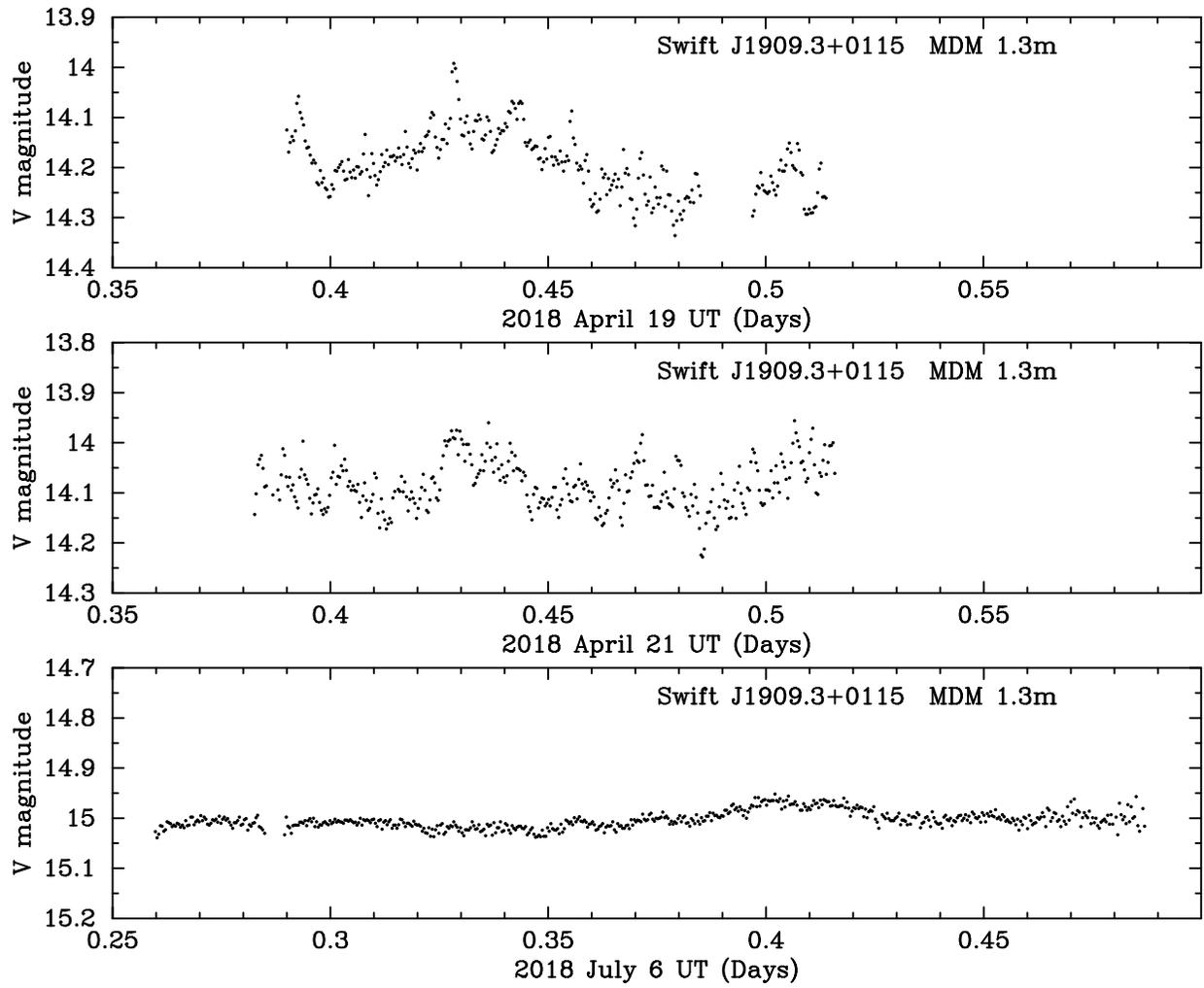}
}
\vspace{-5.4in}   
\caption{
Additional light curves of \swiftOneNineOh\ displaying a wider range of
flux states than in Figure~\ref{fig:swiftj1909}.  Together, these
two figures show that the amplitude of flickering decreases markedly
as the average flux decreases.
}
\label{fig:swiftj1909_more}
\end{figure}

We then obtained $V$-band time-series photometry on ten nights between 2018 March 15
and July 6.  Six of these runs were 3--5 hours long at 33~s cadence, and are shown
in Figures~\ref{fig:swiftj1909} and \ref{fig:swiftj1909_more}.  Flickering with
full amplitude of $\sim0.2$ mag is seen, but no definite coherent period is detected.
Among these nights the mean magnitude ranged from $V=14.1-15.0$.
Figure~\ref{fig:swiftj1909_more} shows the brightest and faintest states
observed which, together with Figure~\ref{fig:swiftj1909}, illustrates
that the amplitude of flickering is correlated with brightness.
Note that the star appeared in its lowest
state on 2018 July~6 (Figure~\ref{fig:swiftj1909_more}), when its
flickering amplitude was at minimum.

We obtained more spectra with OSMOS and modspec through the 2018 observing 
season (see Table \ref{tab:swiftobsjournal}).  The emission-line strength varies dramatically.  The emission equivalent width
of H$\alpha$ falls mostly in the range 15--25\,\AA, but it was as high
as 45\,\AA\ in the discovery spectrum on March 4, and as low as $\sim$2\,\AA\
for some of our June data.  In the latter spectra no emission lines other than H$\alpha$ are detected.  

\begin{figure}
\centerline{
\includegraphics[height=22 cm, trim = 0cm 4cm 1cm 5cm, clip=true]{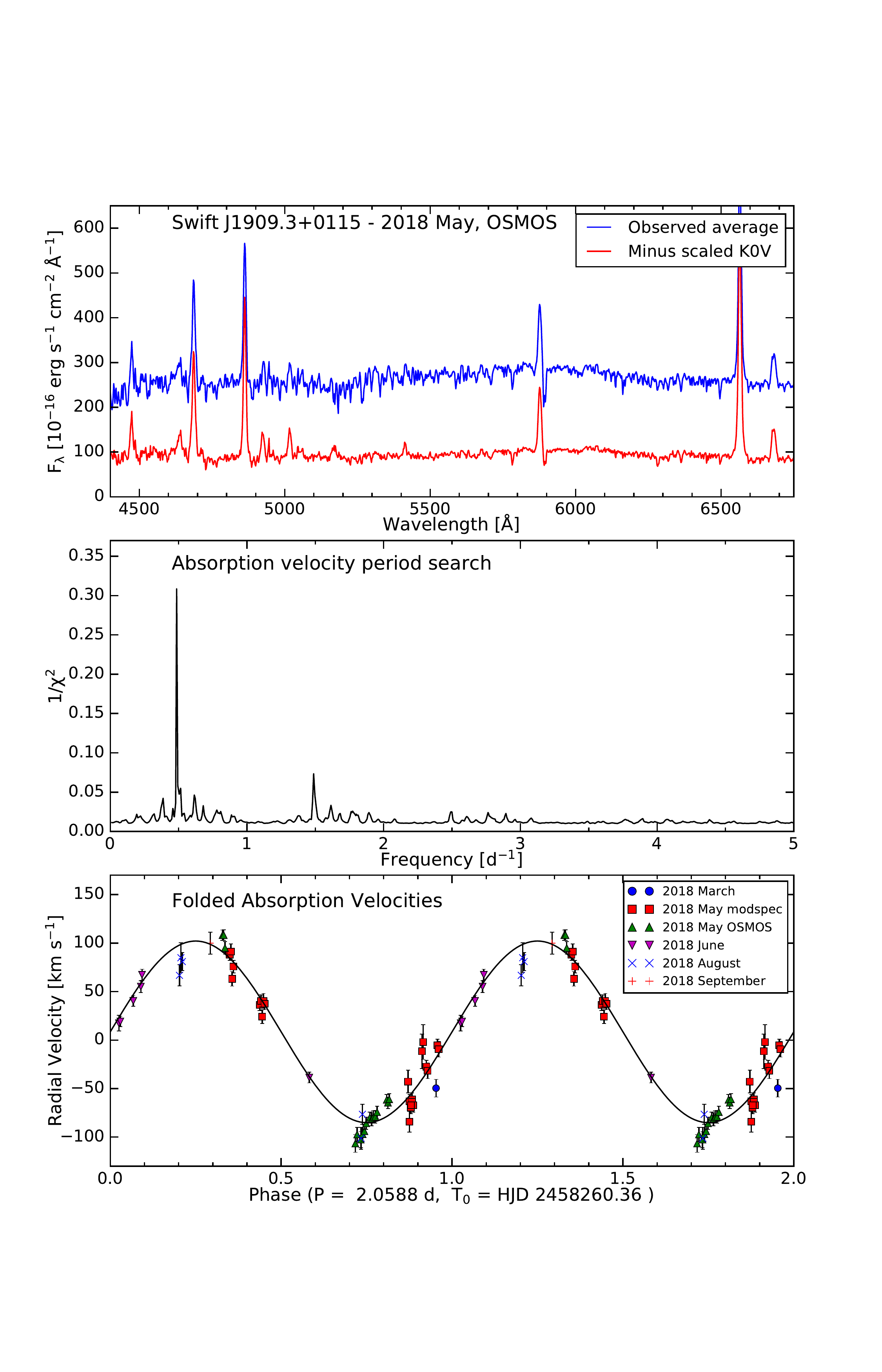}
}
\caption{Top: Spectrum of \swiftOneNineOh. The upper (blue) trace shows the average of the fluxed spectra shifted to the rest frame of the secondary; the lower (red) trace shows the same spectrum after a scaled spectrum of a K0.5~V star has been subtracted.  Middle: Period search of the absorption-line velocities from the 2018 observing season.  The prominent peak is the adopted orbital period. Bottom: Radial velocities of the secondary absorption spectrum folded on the adopted ephemeris, with the best-fitting sinusoid superposed.  All data are plotted a second time for continuity.
}
\label{fig:swift1909_spectra}
\end{figure}

Figure~\ref{fig:swift1909_spectra} (top) shows the mean spectrum from the 
OSMOS data taken 2018 May. It has been corrected for an interstellar
extinction $E(B-V) = 0.4$~mag (see the following discussion).
We characterized the secondary star by averaging
the individual flux-calibrated spectra in the rest frame of the secondary star using the 
radial velocity ephemeris discussed below, and then scaling and subtracting
late-type stellar spectra (also shifted to rest) to obtain the best cancellation 
of the late-type features.  The best match was found using the K0.5~V star
Gliese 567.  We judge the uncertainty in the classification to be about 2 subclasses.
To measure radial velocities of the absorption spectrum, we cross-correlated
the spectra against a composite rest-frame G-K template spectrum, using
the IRAF task {\tt fxcor}, and obtained 52 usable velocities.  
They show a sinusoidal
(almost certainly orbital) modulation near 2.06 days, with
no ambiguity in cycle count over the span of the observations (Figure~\ref{fig:swift1909_spectra},
middle panel).  A sinusoidal fit of the form 
$v(t) = \gamma + K \sin[2\pi (t - T_0) / P]$ gives
$$T_0 = \rm{BJD}\ 2458260.36 \pm 0.02$$
$$P = 2.059 \pm 0.003\ \rm{day}$$
$$K = 94 \pm 5\ \rm{km\ s}^{-1}$$
$$\gamma = 8 \pm 4\ \rm{km\ s}^{-1}.$$
The lower panel of Figure~\ref{fig:swift1909_spectra} shows the folded velocities
and the best fit.

The period is unusually long for a CV, and requires that the companion
be somewhat evolved if it fills its Roche lobe.
The Gaia EDR3 distance is nominally $724\pm11$~pc, and the reddening
maps of \citet{greenreddening} give $E(B-V)=0.42$ at this
distance.  From our spectrophotometry we estimate $V=15.5$ for the
companion star, which implies an absolute magnitude near +5 for the
companion alone, more luminous than a main-sequence star
of its early K spectral type.  Since the faintest state observed
in time series has $V\approx 15$ and very little flickering (Figure~\ref{fig:swiftj1909_more}), it
is possible that accretion makes a smaller contribution to the
light than the companion star in this state.

\subsection{\swiftTwoTwo}
\label{sec:swiftTwoTwo}

\begin{figure}
\centerline{
\includegraphics[width=1.1\linewidth]{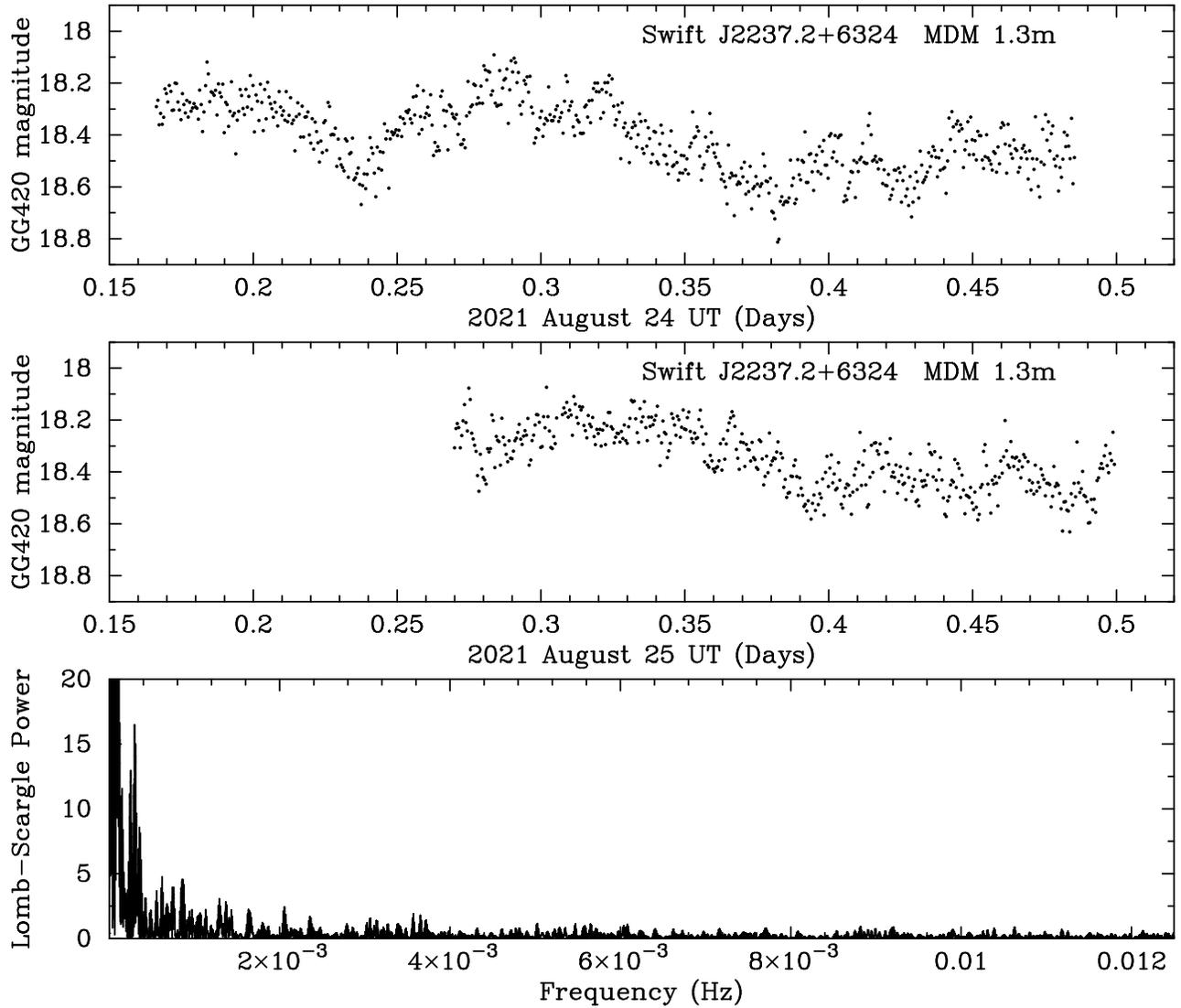}
}
\vspace{-4.in}   
\caption{
Time-series photometry of \swiftTwoTwo\ on two consecutive nights in 2021 August.
Individual exposures are 40~s.  The joint Lomb-Scargle periodogram
shows that there is flickering on a range of timescales, but it does
not confirm the suggested $\sim8$~hr period
from a previous observation (Paper~III).
}
\label{fig:swiftj2237}
\end{figure}

This CV with strong \ion{He}{2} $\lambda4686$ emission
was identified by \cite{lut12b}.  See that paper
for a finding chart.  In Paper~III we showed
a long light curve of \swiftTwoTwo\ from 2016 that has a strong modulation on a 
timescale of $\sim8$~hr, possibly a candidate orbital period.
In 2021 August we obtained two more nights of time-series
photometry that, while displaying flickering on a variety 
of timescales (Figure~\ref{fig:swiftj2237}), do not reproduce
the previously observed light curve.  Thus, the suggested
$\sim8$~hr period is not confirmed.

\section{CONCLUSIONS}
\label{sec:conclusions}

We identified four new CV counterparts of \sw\ BAT survey 
or \int\ sources, and obtained time-resolved
photometry on an additional six that were previously known.

The newly identified \swiftOneNineOh\ has a 2.06~day spectroscopic period, which requires that its optical companion
be slightly evolved to fill its Roche lobe. The companion makes a large contribution to the light.  \swiftOneNineOh\ also displays variable emission-line EWs, and flickering amplitudes that are correlated with continuum brightness, which may be an effect of dilution by the companion's light.

At least four of the targets are IPs based on short periods
detected in their optical photometry.  Two of these, \igr\ and \axj,
display periods in agreement with previously known
X-ray values.  However, \axj\ has additional optical periods whose
interpretation is not obvious.  The third IP is the newly discovered
\pbcOneEight\ with a 311.8~s period as well as a 0.221909~day eclipse
period.  It is not clear if 311.8~s is the spin period or a beat period.
A 152.49~s period is detected from \igrOneEightFour, the fourth IP in this
study.  The previously identified \swiftOhEight\ was reobserved in states
differing by $\approx4$~mag.  In the high states, significantly different
periods were detected, suggesting that they are transient QPOs rather than
spin-related periods. 

Based on their strong \ion{He}{2} emission lines, \swiftOneNineOh\ and \swiftTwoTwo\
warrant further testing for IP signatures.  While evidence of spin is not obvious 
in their optical light curves, spin periods are often more easily detected in X-rays.
This proved to be the case for the 803.5~s period detected in X-rays from \swiftOhSevenOne.
We also recommend \swiftOhEight\ (but in a high state as discussed above)
for X-ray timing, as well as \pbcOneEight\ to clarify the nature of its period structure
around 311~s.

\begin{acknowledgements}

MDM Observatory is operated by Dartmouth College,
Columbia University, the Ohio State University, Ohio University,
and the University of Michigan.  We thank the Tohono O'odham
Nation for leasing the land on which MDM Observatory sits.
This work has made use of data from the European Space Agency (ESA)
mission Gaia (\url{https://www.cosmos.esa.int/gaia}), processed by
the Gaia Data Processing and Analysis Consortium (DPAC,
\url{https://www.cosmos.esa.int/web/gaia/dpac/consortium}). Funding
for the DPAC has been provided by national institutions, in particular
the institutions participating in the Gaia Multilateral Agreement.

\end{acknowledgements}

\clearpage

\movetabledown=2.2in
\begin{rotatetable}
\begin{deluxetable}{lcccccccllclc}
\label{tab:objects}
\tablecolumns{13}
\tabletypesize{\footnotesize}
\tablewidth{0pt}
\tablecaption{Basic Data on Stars Observed}
\tablehead{
\colhead{Name} &
\colhead{R.A.(J2000)\tablenotemark{\rm a}} &
\colhead{Decl.(J2000)\tablenotemark{\rm a}} &
\colhead{$\pi$\tablenotemark{\rm a}} &
\colhead{$g$\tablenotemark{\rm b}} & 
\colhead{$r$\tablenotemark{\rm b}} &
\colhead{$i$\tablenotemark{\rm b}} &
\colhead{Data\tablenotemark{\rm c}} & 
\colhead{Class} &
\colhead{$P_{\rm orb}$} &
\colhead{$P_{\rm spin}$} & 
\colhead{$P_{\rm alt}$} & 
\colhead{Ref\tablenotemark{\rm d}}\\
\colhead{} &
\colhead{(h\ \ \ m\ \ \ s)} &
\colhead{($^\circ$\ \ \ $'$\ \ \ $''$)} & 
\colhead{(mas)} & 
\colhead{} &
\colhead{} & 
\colhead{} &
\colhead{} & 
\colhead{} &
\colhead{(day)} & 
\colhead{(s)} &
\colhead{(s)} &
\colhead{}
}
\startdata
\swiftOhSevenOne\   & 07 17 48.260 & $-21$ 53 01.49 &  0.399(96)   & 18.72 & 18.14 & 17.70 & T &  & 0.2298(10) & $803.5\pm0.7$ & & 1, 2\\
\swiftOhSevenFour\  & 07 49 31.985 & $-32$ 15 36.49 &  0.269(98)   & \dots & \dots  & \dots  & T &  \\
\swiftOhEight\   & 08 20 34.091 & $-28$ 04 58.45 &  \dots          & 17.19 & 17.07 & 16.74 & T    & IP? & & $2353.5\pm3.3$ & $2484(31)$ & 3 \\
                &              &                &                 &       &        &      &      &     & 0.0608(49)  & 1672(60) & & 4 \\
\igrOneEightOh\ & 18 01 12.511 & $-35$ 39 12.24 & 0.607(43)      & \dots & \dots & \dots & I,T & & & & & \\
\igr\            & 18 15 03.847 & $-10$ 51 35.01 & $-0.75\pm1.14$  & 21.76 & 20.09 & 19.24 & T    & IP & & $389.0\pm1.8$ & & 3 \\
                                                                   & & & & & & & & & & $390.5$ & & 5 \\
\axj\            & 18 32 19.304 & $-08$ 40 30.41 & 1.23(36)     & 20.55 & 19.42 & 18.79 & T    & IP & & $1554.2\pm1.7$ & $2303(2)$ & 3 \\
                                                                   & & & & & & & & & & $1549.1\pm0.4$ & & 6 \\
                                                                   & & & & & & & & & & $1552.3^{+2.3}_{-0.8}$ & & 7 \\
\pbcOneEight\    & 18 41 04.160 & +01 37 55.76 & 1.49(30)      & 21.64 & 19.73 & 18.48 & I,T   & IP & 0.2219088(11) & 311.805(4) & & 3 \\ 
\igrOneEightFour\ & 18 43 11.424 & $-05$ 05 45.56 & 0.44(15)   & 19.08 & 18.64 & 18.18 & I,T   & IP &  & 152.49(2) & & 3 \\
\swiftOneNineOh\ & 19 09 20.906 & +01 12 24.56 & 1.381(21)     & 15.12 & 14.64 & 13.99 & I,S,T &   & 2.059(3) & & & 3 \\
\swiftTwoTwo\   & 22 36 37.401 & +63 29 33.60 & 0.98(15)      & 20.07 & 18.98 & 18.26 & T & & & & & \\
\enddata
\tablenotetext{\rm a}{Coordinates and parallax are from the Gaia extended third data release (EDR3; \citealt{bro21}), which are corrected for proper motion to epoch 2016.0.}
\tablenotetext{\rm b}{Mean PSF magnitudes from Pan-STARRS.}
\tablenotetext{\rm c}{Types of data presented here:
I---optical spectroscopic identification;
S---time-resolved spectroscopy;
T---time-series photometry.}
\tablenotetext{\rm d}{Reference for period(s):
(1) \citealt{paperII}, optical; (2) this paper, X-ray; (3) this paper, optical; (4) \citealt{nuc20}, X-ray; (5) \citealt{wor20}, X-ray; (6) \citealt{sug00}, X-ray; (7) \citealt{kau10}, X-ray.}
\end{deluxetable}
\end{rotatetable}

\clearpage

\begin{deluxetable}{cc}
\label{tab:eclipse}
\tablecolumns{2}
%\tablewidth{200pt}
\tablecaption{Eclipse Timings of \pbcOneEight}
\tablehead{
\colhead{Cycle} &
\colhead{Mid-Eclipse} \\
\colhead{} &
\colhead{(BJD)}
}
\startdata
0     &  2458630.88975 \\
4     &  2458631.77730 \\
9     &  2458632.88707 \\
144   &  2458662.84481 \\
148   &  2458663.73211
\enddata
\end{deluxetable}

\begin{deluxetable}{ccrrrrc}
\label{tab:swiftobsjournal}
\tablecolumns{7}
\tablecaption{Spectroscopy Journal for \swiftOneNineOh}
\tablehead{
\colhead{Start} &
\colhead{Tel., Instr.\tablenotemark{a}} &
\colhead{HA start} &
\colhead{HA end} &
\colhead{Exp} &
\colhead{$N_{\rm exp}$} &
\colhead{H$\alpha$ EW\tablenotemark{b}} \\
\colhead{(UT)} & 
\colhead{} &
\colhead{(hh:mm)} &
\colhead{(hh:mm)} &
\colhead{(s)} &
\colhead{} &
\colhead{(\AA)}
} 
\startdata
2018-03-04 12:48 & 2.4-m, O & $-$02:59 & $-$02:49 &  600 &  1 & 45 \\
2018-05-09 07:45 & 1.3-m, M & $-$03:42 & $-$02:47 &  600 &  5 & 17 \\
2018-05-10 09:43 & 1.3-m, M & $-$01:41 & $-$00:48 &  600 &  5 & 15 \\
2018-05-11 07:22 & 1.3-m, M & $-$03:58 & +00:25 &  600 &  6 & 15 \\
2018-05-13 08:33 & 1.3-m, M & $-$02:39 & +00:21 &  720 &  5 & 14 \\
2018-05-14 08:11 & 1.3-m, M & $-$02:57 & $-$02:15 &  600 &  4 & 14 \\
2018-05-20 11:22 & 2.4-m, O & +00:37 & +01:09 &  600 &  3 & 27 \\
2018-05-21 06:31 & 2.4-m, O & $-$04:10 & +00:57 &  900 & 14 & 26 \\
2018-06-05 07:37 & 2.4-m, M & $-$02:04 & +01:28 &  600 &  4 & 3.6 \\
2018-06-06 11:08 & 2.4-m, M & +01:31 & +01:43 &  720 &  1 & 10 \\
2018-06-07 11:08 & 2.4-m, M & +01:35 & +01:45 &  600 &  1 & 1.9 \\
2018-08-31 03:37 & 1.3-m, M & $-$00:22 & +00:15 &  720 &  3 & 18 \\
2018-09-01 05:52 & 1.3-m, M & +01:57 & +02:21 &  720 &  2 & 10 \\
2018-09-27 02:28 & 2.4-m, M & +00:15 & +00:25 &  600 &  1 & 34\\
\enddata
\tablenotetext{a}{M = modspec; O = OSMOS.}
\tablenotetext{b}{Equivalent width of H$\alpha$, positive for emission.}
\end{deluxetable}

\end{document}